\documentclass[10pt,preprint]{aastex}

\usepackage{graphicx}
\newcommand{\esp}{\mathsf{E}}
\newcommand{\var}{\mathsf{var}}
\newcommand{\cov}{\mathsf{cov}}

\shorttitle{Dynamic range in coronagraphic images}
\shortauthors{R. Soummer etal.}

\begin{document}

\title{Speckle noise and dynamic range in coronagraphic images}
\slugcomment{Accepted for publication in ApJ}

\author{R\'emi Soummer \altaffilmark{1}}
\affil{American Museum of Natural History, 79th St. at Central Park West, New York, NY 10024, USA}
\email{rsoummer@amnh.org}

\and
\author{Andr\'e Ferrari, Claude Aime}
\affil{LUAN, Universit\'e de Nice Sophia Antipolis, Parc Valrose, 06108 Nice, France}
\email{ferrari@unice.fr, aime@unice.fr}

\and
\author{Laurent Jolissaint}
\affil{Herzberg Institute of Astrophysics, 5071 West Saanich Road, Victoria, B.C. V9E 2E7, Canada}
\email{laurent.jolissaint@nrc-cnrc.gc.ca}

\altaffiltext{1}{Michelson Fellow, under contract with the Jet Propulsion Laboratory (JPL) funded by NASA through the Michelson Fellowship Program. JPL is managed for NASA by the California Institute of Technology.}

\begin{abstract}
This paper is concerned with the theoretical properties of high contrast coronagraphic images in the context of exoplanet searches. 
We derive and analyze the statistical properties of the residual starlight in coronagraphic images, and describe the effect of a coronagraph on the speckle and photon noise. 
Current observations with coronagraphic instruments have shown that the main limitations to high contrast imaging are due to residual quasi-static speckles. 
We tackle this problem in this paper, and propose a generalization of our statistical model to include the description of static, quasi-static and fast residual atmospheric speckles. 
The results provide insight into the effects on the dynamic range of wavefront control, coronagraphy, active speckle reduction, and differential speckle calibration. The study is focused on ground-based imaging with extreme adaptive optics, but the approach is general enough to be applicable to space, with different parameters. 
\end{abstract}

\keywords{instrumentation: adaptive optics, instrumentation: high angular resolution}

\section{Introduction}
\label{sect:intro} 

Direct imaging of faint companions or planets around a bright star is a very difficult task, where the contrast ratio and the angular separation are the observable parameters. The problem consists of detecting a faint source over a bright and noisy background, mainly due to the diffracted stellar light. High contrast ratios and small angular separations correspond to the most difficult case. Typically, for extra-solar giant planets, contrast ratios of about $10^{-7}$ are expected in the near infrared (J,H,K bands), based on models for relatively young objects of about 100 Myr \citep{CB00,BCB03,BSH04}. According to these models, older objects would be an order of magnitude fainter. Terrestrial planets are much fainter than giant planets, about 3 to 4 orders magnitudes fainter depending on the wavelength range.

\begin{figure}[htbp]
\center
\resizebox{.4\hsize}{!}{\includegraphics{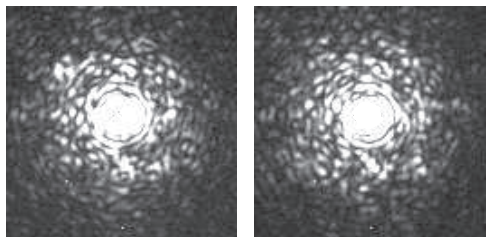}}
 \caption{Direct images of the bright star HD 137704 (magnitude V=5.47) obtained with the Adaptive Optics system at the Palomar Hale telescope on June 6th 2004. 
 The core of the image has been saturated to illustrate the speckles. The bright speckles localized at the position of the diffraction rings are called pinned speckles. Their statistical properties in direct and coronagraphic images are discussed at length in this paper.}
 \label{FigPalomar}
\end{figure}

In the case of ground-based observations with Adaptive Optics (AO), the residual uncorrected aberrations produce random intensity fluctuations of the background, which appear as speckles in the field. In direct non-coronagraphic high quality images, these speckles mainly appear at the position of the diffraction rings of the star. In Fig.\ref{FigPalomar}, we show two high-quality point spread functions (PSF) obtained with the AO system at Palomar \citep{TDB00,HBP01}, where the speckles are clearly visible. This phenomenon, also known as ``speckle pinning'' \citep{BDT01}, can be explained using an expansion of the point spread function (PSF) \citep{SLH02,SHM03,B03,PSM03,Blo04}. 
An alternative approach using a statistical model \citep{AS04} enables a deeper understanding of the phenomenon, especially from the statistical point of view, by providing information on the speckle variance. The effect of a coronagraph on speckle noise is well explained with this approach. In particular, we show that static aberrations produce residual speckle pinning after a coronagraph, which has important implications in high contrast imaging. Statistical properties of long exposure AO images were also studied by \citet{FC04}. 

The dynamic range of a coronagraphic image corresponds to the faintest companion that can be detected at a given position in the field, at the detection limit. It is usually expressed as a magnitude difference or an intensity ratio, relative to the unocculted central star at some signal to noise ratio (SNR) level. Although the dynamic range is a two-dimensional map of the detection sensitivity in the field, it is often represented as a radial profile showing the magnitude difference $\Delta m$ as a function of the angular separation $r$. A radial profile should be acceptable for most applications, especially when the coronagraph does not have particular asymmetries \citep{L39,ASF02,S05}. Otherwise, a two-dimensional map of sensitivity is required when the coronagraph shows an asymmetric response \citep{KS03,KVS03,RRB00}.
Understanding, measuring and predicting the dynamic range is still one of the important issues in this field, with implications for instrumentation (design, observing strategies) and data reduction and analysis.

During the design phase of future planet finder instruments using Extreme Adaptive Optics (ExAO) and coronagraphy, for example the Gemini Planet Imager \citep{MBW04,MGP06}, or the ESO/VLT SPHERE \citep{BMD05,FRM05}, it is necessary to anticipate what part of the observable parameter-space (contrast \textit{vs.} separation) can be probed, and link it to the actual physical parameter space (mass \textit{vs.} semi major axis). Such studies have been carried out by \citet{Brown04a,Brown04b, Brown05} in the context of terrestrial planet searches. 

When operating an existing high-contrast instrument, like the Lyot project coronagraph \citep{ODN04}, the dynamic range has to be measured to evaluate the performance \citep{SOH06,HOS06}. In the case of a detected object, photometry and astrometry \citep{DHO06,SO06,MLM06} are necessary to help determine the objects characteristics. Dynamic range computations are also important in the case of non-detection, to determine which part of the parameter space has been probed by the experiment, and which physical objects can be ruled out. 

The dynamic range is limited by the intensity fluctuations close to the star. These are due to several sources: speckle noise, photon noise, detector noise, background noise etc. Speckle noise is known to be the main limitation for high contrast imaging, either in direct images \citep{MDN03,MDN05,MMH05,MLD06}, or coronagraphic images \citep{BML97,OGK01,BCL03,BRB04,HOS06}. Speckles find their origin in wavefront imperfections (amplitude and phase errors), whether they correspond to uncorrected atmospheric residual errors (residual atmospheric speckles) or slow-varying wavefront caused by mechanical or thermal deformations. The main problem comes from these quasi-static speckles, which can be calibrated either using active pre-focal methods \cite{MYS95,GKV06,BT06} or using post-processing \citep{MDR00,SF02,MLD06}. An alternative approach based on non-redundant masking has been achieved by \citet{LMI06}.
Assuming a good enough calibration of these quasi-static speckles, the physical limit of the system is set by the residual atmospheric aberrations.
The noise limitation in high dynamic range images has been studied by several authors, using numerical simulations \citep{B04,CBB06}, or other theoretical or empirical approaches \citep{A94,RWN99,PSM03,G05, SSS05}.

In \cite{AS04}, we modeled the statistics of AO-corrected, direct images, and discussed the effect in terms of signal to noise ratio on these images qualitatively. In this paper, we examine the effects a coronagraph has on the properties of residual speckles. A semi-analytical method to compute the dynamic range can be derived from the statistical properties of the speckle and photon noise. We compare our results with purely numerical simulations. Results in this paper are presented using an Apodized Pupil Lyot Coronagraph (APLC) as an example \citep{ASF02,SAF03,S05}, but are valid for any other type of mask coronagraph \citep{RRB00,KT02,SDA03,MRA05}, pure apodizers or shaped pupils \citep{JR64,NP01,SAFF03,KVS03,Aime05a}. Furthermore, the statistical model can be modified to include both static and quasi-static aberrations, and we discuss the coherent interaction between residual atmospheric and quasi-static aberrations.

Our theoretical model and results apply for both space and ground-based imaging. However, we illustrate the results with simulations in the case of ground-based Extreme Adaptive Optics (ExAO) and coronagraphy.

\section{Propagation through a coronagraph in the presence of aberrations}\label{Sec:Prop}

In this section we assume that all the static aberrations, mainly optical quality (polishing for example) and misalignment errors, can be represented at the entrance pupil of the telescope. The case of out of pupil aberrations is discussed by \citet{MPM06} and does not affect our monochromatic statistical model. 
We consider an instrument with an ExAO system and a generic coronagraph which can describe any type of mask designs \citep{RRB00,ASF02,KT02,SDA03,S05,MRA05}. The formalism also applies to the shaped pupil approach which corresponds to the case of direct imaging with apodization \citep{KVS03}. The ExAO and telescope characteristics are chosen to be consistent with current or future projects on eight-meter telescopes and illustrations are given for an APLC.

These coronagraphs consist of optical filtering in four successive planes denoted 1,2,3,4 hereafter. The first plane corresponds to the entrance aperture (possibly apodized), the second plane is the focal plane where a mask is applied (opaque, graded or phase-shifting), the third plane corresponds to a relay pupil plane where a diaphragm is applied (the Lyot Stop), and finally the fourth plane corresponds to the final focal plane. 

\begin{figure}[htbp]
\center
\resizebox{0.8\hsize}{!}{\includegraphics{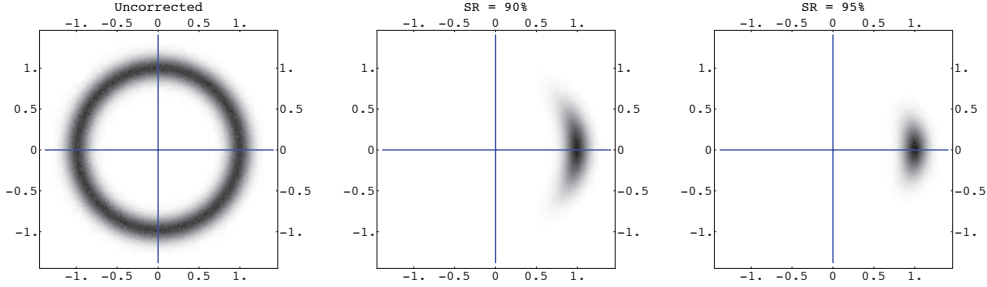}}
 \caption{Probability Density Functions (PDF) of the pupil plane complex amplitude. Left: PDF of uncorrected atmospheric wavefronts obtained from a von Karman power spectrum model, with an outer scale $L_0=20m$ and a seeing $\omega_0=0.8$ arcsec. The phase excursion is uniform over $(0-2\pi)$ and the thickness of the annulus corresponds to the amplitude scintillation. Center and Right: PDFs of a AO-corrected wavefronts for Strehl Ratio of $90\%$ (center) and $95\%$ (right). These distributions in the complex plane look like de-centered crescents. The length of the crescent corresponds to the phase excursion, and the thickness to scintillation, which is assumed to have no effect on the AO system. The effect of the improved AO correction on the phase excursion is obvious between these two figures. }
 \label{fig1PDF}
\end{figure}

In the pupil plane, we model the wavefront using a static phase aberration $\varphi_s(\textbf{r})$, a residual random atmospheric phase $\varphi(\textbf{r})$, and amplitude aberrations $\rho(\textbf{r})$, including the eventual apodization. The complex amplitude is:
\begin{equation}\label{Eq0}
\Psi_1(\textbf{r})=P(\textbf{r}) \,\rho(\textbf{r}) e^{j (\varphi(\textbf{r})+\varphi_s(\textbf{r}))},
\end{equation}
where the function $P(\textbf{r})$ describes the normalized aperture transmission: $\int P(\textbf{r}) d\textbf{r}=1$, and $\textbf{r}=(x,y)$ is the coordinate vector, used in both pupil and field. 
Following the notations of \citet{AS04}, we can write the wavefront complex amplitude at the entrance pupil as the coherent sum of three terms:
\begin{equation}\label{Eq1}
\Psi_1(\textbf{r})=\left[A+A_s(\textbf{r})+a(\textbf{r})\right]\,P(\textbf{r}),
\end{equation}
where $A$ is a deterministic term corresponding to a perfect plane wave, $A_s(\textbf{r})$ is a deterministic complex term corresponding to the static aberrations, and $a(\textbf{r})$ is a random term with zero mean $(\esp[a(\textbf{r})]=0)$ corresponding to the uncorrected part of the wavefront.
The probability density function (PDF) of this complex amplitude is illustrated in Fig.\ref{fig1PDF} without and with AO. Static aberrations are not included in this figure.
$A$ is defined as the mean of the complex amplitude, averaged over the pupil: 
\begin{equation}\label{Eq:DefA}
A=\esp[\int\Psi_1(\textbf{r})P(\textbf{r})d\textbf{r}].
\end{equation}
$|A|^2$ is therefore the Strehl Ratio of the system \citep{Hardy98}.\\
With $\esp[a(\textbf{r})]=0$, Eq.\ref{Eq1} implies $\esp[\Psi_1(\textbf{r})]=A+A_s(\textbf{r})$. Integrating this equation, we obtain:
\begin{equation}
\esp[\int \Psi_1(\textbf{r})P(\textbf{r})d\textbf{r}]=A+\int A_s(\textbf{r})P(\textbf{r})d\textbf{r},
\end{equation}
and therefore:
\begin{equation}\label{intAsP}
\int A_s(\textbf{r})P(\textbf{r})d\textbf{r}=0.
\end{equation}
Assuming that the phase errors are stationary over the aperture, we obtain:
\begin{equation}\label{intAP}
\int a(\textbf{r})P(\textbf{r})\approx 0.
\end{equation} 
Note that the difference between Eq.\ref{intAsP} and Eq.\ref{intAP} comes from the fact that $A_s(\textbf{r})$ is deterministic and $a(\textbf{r})$ is random: Although $A_s(\textbf{r})$ can be defined with zero-mean over the aperture, each independent realizations of $a(\textbf{r})$ do not necessarily have an exactly zero average over the aperture. 

\begin{figure}[htbp]
\center
\resizebox{.4\hsize}{!}{\includegraphics{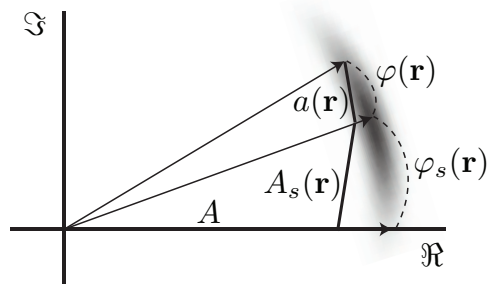}}
 \caption{Illustration of the decomposition of the wavefront as the sum of three complex vectors. We consider a static phase term $\varphi_s$ and a residual atmospheric phase $\varphi$. $A$ is the mean wavefront and $|A|^2$ is therefore the Strehl Ratio. $A_s$ corresponds to the static aberrations, and $a$ to the zero-mean error term.}
 \label{Fig:Vectors}
\end{figure}

\begin{figure}[htbp]
\center
\resizebox{.5\hsize}{!}{\includegraphics{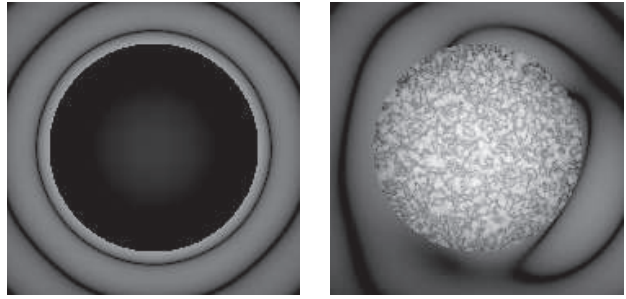}}
 \caption{Example of propagation though a coronagraph for a monochromatic APLC, corresponding to Eq.\ref{Eq5}. Left: intensity in the Lyot stop plane in the perfect case $(|\Psi_{c}(\textbf{r})|^2)$ . Most of the light is rejected outside of the pupil aperture and will be eliminated by the Lyot Stop. Right: intensity of the speckle term $|\Psi_a(\textbf{r})|^2$, assumed alone. Most of the energy remains inside the geometric aperture and will not be eliminated by the Lyot Stop and appear as residual speckles in the final image. Both images are represented with the same scale.}
 \label{fig2}
\end{figure}

\begin{figure}[htbp]
\center
\resizebox{.5\hsize}{!}{\includegraphics{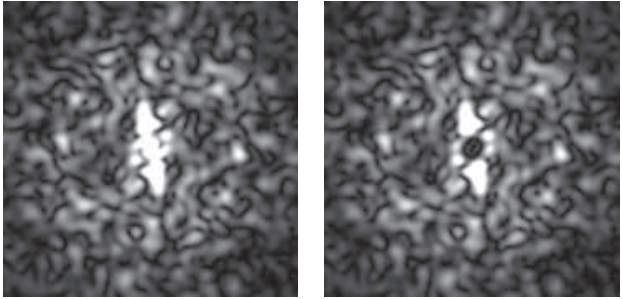}}
 \caption{Illustration of the terms of Eq.\ref{Eq6}. Left: modulus of the speckle term alone $S(\textbf{r})=\mathcal{F}[a(\textbf{r})\, P(\textbf{r})]$. Right: actual term $\Psi_4(\textbf{r})$ represented in modulus at the same scale. The cross terms of Eq.\ref{Eq6} only affect the central part of the image, inside an area corresponding to the occulting mask size.}
 \label{figSpeckCrossTerm}
\end{figure}

The specific case of quasi-static aberrations is not considered in this section and will be treated in Sec.\ref{SecQuasiStatic}. The three terms of Eq.\ref{Eq1} are illustrated in Fig.\ref{Fig:Vectors}.
The length of the vector $A$ is arbitrary in the figure, to illustrate that the modulus of $A$ is not unity and that the vectors $A$, $A_s$, and $a$ are defined according to the definitions above.
In the first focal plane, a coronagraphic mask is applied at the center of the image of the star. Writing the mask transmission as $1-M(\textbf{r})$, allows us to accommodate any type of mask coronagraph, including Lyot, APLC, Band-Limited, Phase Masks. For example, a classical hard-edged Lyot coronagraph (or APLC), is described using a top-hat function for $M$. The complex amplitude of the wave in the focal plane is given by a scaled Fourier Transform (FT) of this pupil amplitude \citep{Goo96}:
\begin{equation}\label{Eq2}
\Psi_2(\textbf{r})=\mathcal{F}[\Psi_{1}(\textbf{r})] \left(1-M(\textbf{r})\right),
\end{equation}
where the symbol $\mathcal{F}$ denotes the scaled FT. For clarity, we will omit the wavelength-dependent scaling factors. For the complete chromatic formalism see for example \citet{ASF02,SAF03,aime05b}.
In the next pupil plane, the complex amplitude before the Lyot stop $P'(\textbf{r})$ is also the sum of three terms:
\begin{equation}\label{Eq5}
\Psi_3(\textbf{r})=A \, \Psi_{c}(\textbf{r}) + \Psi_s(\textbf{r})+\Psi_a(\textbf{r}),
\end{equation}
where $\Psi_{c}(\textbf{r})= P(\textbf{r})- P(\textbf{r})\ast \mathcal{F}[M(\textbf{r})]$ is the complex amplitude in the lyot stop plane for a perfect wavefront \citep{SAF03}. The two other terms $ \Psi_s(\textbf{r})$ and $\Psi_a(\textbf{r})$ correspond to the propagation of the terms $A_s(\textbf{r})$ and $a(\textbf{r})$ respectively. For example: 
\begin{equation}\label{Eq:Psia}
 \Psi_a(\textbf{r})=a(\textbf{r})P(\textbf{r})- a(\textbf{r})P(\textbf{r})\ast \mathcal{F}[M(\textbf{r})].
 \end{equation}
The perfect coronagraph term $\Psi_{c}(\textbf{r})$ and the term $\Psi_a(\textbf{r})$ are shown in intensity in Fig.\ref{fig2}. The coronagraphs rejects most of the starlight outside the image of the aperture in the Lyot plane for the perfect part of the wave, but most of the energy remains inside the aperture for the speckle part.
The Lyot stop is applied in this plane. In the case of an APLC, the Lyot stop is identical to the entrance pupil; in all other cases, the Lyot stop is undersized. With $P(\textbf{r})P'(\textbf{r})=P'(\textbf{r})$, and with the notations $S(\textbf{r})=\mathcal{F}[a(\textbf{r})\, P'(\textbf{r})]$ and $S_s(\textbf{r})=\mathcal{F}[A_s(\textbf{r})\, P'(\textbf{r})]$, we obtain the complex amplitude in the final focal plane: 
\begin{eqnarray}\label{Eq6}
\Psi_4(\textbf{r})&=&A\, \Psi_{d}(\textbf{r})\nonumber\\
&&+S_s(\textbf{r})-(S_s(\textbf{r}) M(\textbf{r}))\ast \mathcal{F}[P'(\textbf{r})]\nonumber\\
&&+S(\textbf{r})-(S(\textbf{r})M(\textbf{r}))\ast \mathcal{F}[P'(\textbf{r})],
\end{eqnarray}
where $\Psi_{d}$ denotes the focal wave amplitude of the coronagraph in the perfect case, following the notations of \citet{ASF02}. 
The convolution product $(S(\textbf{r})M(\textbf{r}))\ast \mathcal{F}[P'(\textbf{r})]$ in Eq.\ref{Eq6} has a negligible effect outside the mask area. Indeed, the spatial extension of $(S(\textbf{r})M(\textbf{r}))$ is limited to the occulting mask area, and $\mathcal{F}[P'(\textbf{r})]$ is a rapidly decreasing function (the Airy amplitude in a perfect case) whose characteristic size is $\lambda/D$. The result of the convolution does not extend much beyond the mask area, which is illustrated in Fig.\ref{figSpeckCrossTerm}, where we show an example of the effect of the convolution term using a numerical simulation. This can also be explained by considering the propagation of phase ripples through a coronagraph and constructing a Bode diagram \citep{SSO07}.

Grouping all the deterministic terms together, we introduce
\begin{equation}\label{Eq7}
\tilde{C}(\textbf{r})=A\; \Psi_{d}(\textbf{r})+S_s(\textbf{r})-(S_s(\textbf{r}) M(\textbf{r}))\ast \mathcal{F}[P'(\textbf{r})],
\end{equation}
and we obtain a similar expression to the case without coronagraph where $C(\textbf{r})=A\;\mathcal{F}[P(\textbf{r})]$ \citep{AS04}:
\begin{equation}\label{psi4}
\Psi_{4}(\textbf{r}) = \tilde{C}(\textbf{r})+S(\textbf{r}).
\end{equation}
The wave amplitude after a coronagraph appears as a sum of a deterministic term $\tilde{C}(\textbf{r})$, and a random term $S(\textbf{r})$, at each position in the focal plane, outside the mask area.
The deterministic term $\tilde{C}(\textbf{r})$ corresponds to the focal plane complex amplitude of the coronagraph in the presence of static aberrations. 
The random term $S(\textbf{r})$, associated with the speckles, is identical to the case without coronagraph \citep{AS04}: the coronagraph has a negligible effect on the random part of the wavefront, as illustrated in Fig.\ref{fig2} and Fig.\ref{figSpeckCrossTerm}.
Formally, the effect of the coronagraph is to replace the wave amplitude without coronagraph $C(\textbf{r})$ by the coronagraphic amplitude $\tilde{C}(\textbf{r})$. In the case of pure apodizers (shaped pupils), the direct apodized term $C(\textbf{r})$ is used.

The random term $S(\textbf{r})$ is non stationary in the field. The profile for $S(\textbf{r})$ can be computed from a simulation of the AO system, as we detail in Sec.\ref{Sec:NumSims}. Low-order aberrations can also be included in this description, but usually require a specific study, as for example in \citet{SSS05,SSO07}.
The profile or the two-dimensional map for $\tilde{C}(\textbf{r})$ can be computed independently, considering a perfect coronagraph in the presence of deterministic static aberrations (and normalized to the SR). 
Even in the case of an ideal coronagraph that cancels all the star light for a perfect wave, the deterministic term $\tilde{C}(\textbf{r})$ still contains the terms due to the static aberrations and will contribute to speckle pinning, as discussed below.

\section{Statistical properties of direct or coronagraphic images}

\subsection{Statistical model and properties of speckles}

\subsubsection{Complex amplitude distribution}\label{ComplexDistrib}

In this section, we discuss the distribution of the complex amplitude in the focal plane. We consider the case of monochromatic direct images for simplicity. The case of coronagraphic images is formally identical to the coronagraphic case, according to the approximations described in the previous paragraph (Eq.\ref{psi4}).
The focal plane complex amplitude is the Fourier transform of the pupil plane complex amplitude:
\begin{equation}\label{EqPsi2bis}
\Psi_2(\textbf{r})=\int P(\textbf{u}) (A+a(\textbf{u}))\, e^{-2\imath \pi \textbf{u}.\textbf{r}} \mathrm{d}\textbf{u}.
\end{equation}
The complex amplitude in the focal plane is therefore a sum of the random complex term $a(\textbf{r})$ weighted by the Fourier complex phasors. At the center of the image, the Fourier phasors vanish, so a special treatment for this particular case (and the transition region around it) is necessary \citep{SF07}.
Outside the central point of the image, the distribution of the complex amplitude can be derived using known results in Signal Processing, based on reasonable assumptions. We assume that the complex amplitude in the pupil plane can be represented by discrete values (an implicit assumption in any numerical simulation), and that the correlation of the complex amplitude between two points in the pupil plane decreases with distance between them. Under these hypotheses, it can be shown that the distribution of the complex amplitude in the focal plane is asymptotically circular Gaussian \citep{Bri81}. We remind the reader here that if the real and imaginary parts of a complex number $z$ are Gaussian, its distribution is said to be Gaussian. If the real and imaginary parts are independent and have same variance, the distribution is said to be circular Gaussian and denoted $z \sim \mathcal{N}_c(0,\sigma^2)$. See Fig.2 of \citet{AS04} for an illustration of the focal plane PDFs. 
The circularity of the Gaussian distribution is due to the Fourier phasors mixing the complex amplitudes in the complex plane in the Fourier integral (\ref{EqPsi2bis}), where $u$ varies between $-D/2$ and $D/2$. For positions $r$ in the focal plane such as $r>\lambda/D$, the Fourier phase term therefore varies between 0 and $2\pi$ and this circularization occurs. 
The complex amplitude of the wave in the focal plane $\Psi_4(\textbf{r})$ follows a circular Gaussian law, de-centered by the mean of the amplitude $\tilde{C}(\textbf{r})$ and denoted: $\Psi_4(\textbf{r}) \sim \mathcal{N}_c(\tilde{C}(\textbf{r}),\esp[|S(\textbf{r})|^2])$.

\begin{figure}[htbp]
\center
\resizebox{\hsize}{!}{\includegraphics{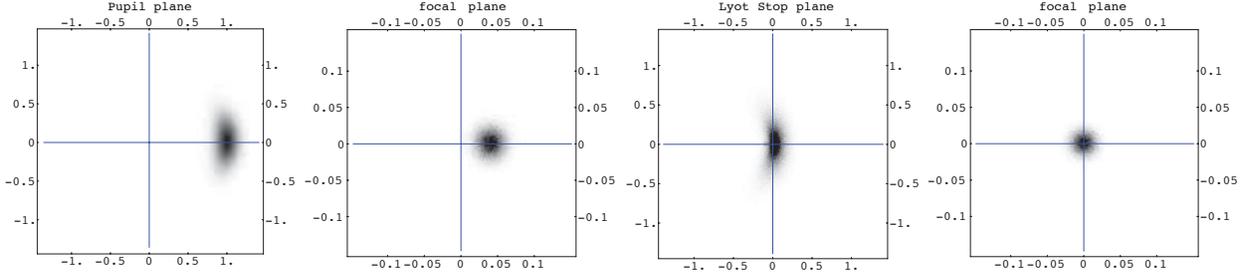}}
 \caption{Complex probability distributions in the four successive coronagraphic planes (pupil, focal, pupil, focal), at an arbitrary angular position in the field $(r=2.6\lambda/D)$. This simulation corresponds to the case of an APLC without static aberrations. The distribution in the pupil plane corresponds to a typical ExAO for an 8-m class telescope delivering 90\% SR, and including scintillation effects. In the focal plane, the distribution at a given position is a decentered Gaussian. In the Lyot stop plane after and APLC, the coronagraph has removed the deterministic part of the entrance pupil wavefront. In the final focal plane, the distribution is close to circular Gaussian distribution. }
 \label{figPDFcorono}
\end{figure}

In Fig.\ref{figPDFcorono}, we give an illustration of the distribution of the complex amplitude in the four successive planes of the coronagraph. This illustration is based on numerical simulations of a perfect APLC coronagraph and of an ExAO system. In the first pupil plane, we have a de-centered crescent (see Fig.\ref{fig1PDF}). In the first focal plane, in this example at the top of an Airy ring, ${C}(\textbf{r})$ has a high absolute value, and the distribution is Gaussian, de-centered by this amount. Detailed illustrations of the decentered Gaussian statistics as a function of the position in the field can be found in \citet{AS04}. In the following Lyot plane the coronagraph almost completely removes the perfect part of the wave (see Fig.\ref{fig2}), and the resulting distribution is similar to the initial distribution of the complex amplitude in the pupil, but centered at the origin.
Finally in the last focal plane, without static aberrations, $\tilde{C}(\textbf{r})\simeq 0$, as $\Psi_d\simeq0$ and the result is a centered circular Gaussian distribution. 
\subsubsection{Intensity distribution}\label{Sec:IntensityDistrib}

In this section we derive the Probability Density Function (PDF) of the intensity from that of the wave complex amplitude. Our problem is formally equivalent to the case of laser speckles added to a coherent background, which has been studied extensively \citep{Goo75,Goo06}, in particular in the context of holography. 
We introduce the two intensity terms:
\begin{eqnarray}
I_c&=&\,|\tilde{C}(\textbf{r})|^2\nonumber\\
I_s&=&\esp[|S(\textbf{r})|^2].
\end{eqnarray}
Note that $I_c$ and $I_s$ are both functions of $\textbf{r}$, and that $I_c$ can describe both the direct or coronagraphic case, with and without static aberrations.
Following Goodman, the joint PDF for the intensity and phase can be obtained from PDF of the complex amplitude, using the simple cartesian-polar change of variables 
$(\eta=\sqrt{I} \cos[\theta],\xi=\sqrt{I} \sin[\theta])$, where the modulus of the Jacobian of this transformation is $\|J\|=1/2$ and integrating the phase $\theta$ to find the PDF for the intensity.

\begin{figure}[htbp]
\center
\resizebox{0.7\hsize}{!}{\includegraphics{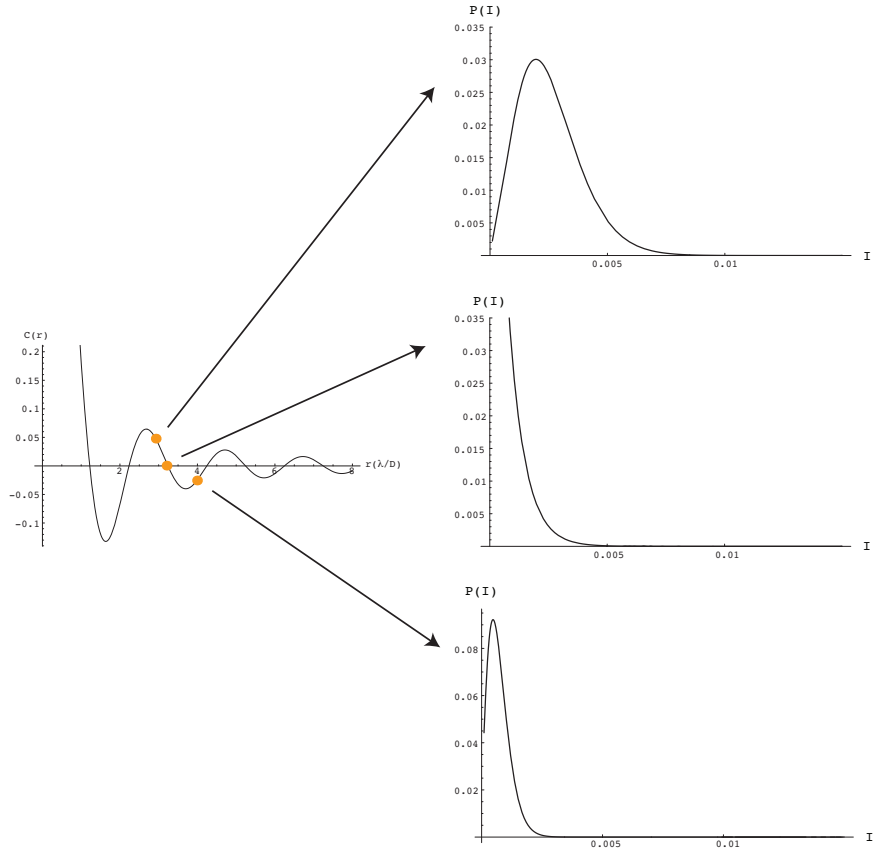}}
 \caption{Probability Density Function of the light intensity at 3 different positions in the focal plane, corresponding to different amplitude $C(\textbf{r})$ (or intensity levels $I_c$). The width of the distribution clearly increases with an increase in the level of the constant intensity background. This approach provides an alternative explanation of speckle pinning, where the constant background corresponding to the perfect part of the wave amplifies speckle fluctuations. }
 \label{figrician}
\end{figure}

An alternative derivation of the PDF for the intensity is to consider the properties of Gaussian distributions. 
As discussed in the previous section, the speckle term $S(\textbf{r})$ is a circular Gaussian distribution $S(\textbf{r})\sim \mathcal{N}_c(0,I_s)$. 
The instantaneous intensity corresponding to the complex amplitude of Eq.\ref{psi4} is simply: 
\begin{eqnarray}\label{Eq:InstantaneousInt}
I &=& |S(\textbf{r})+\tilde{C}(\textbf{r})|^2\nonumber\\
 &=&\left(\Re{[\tilde{C}(\textbf{r})+S(\textbf{r})]}\right)^2+\left(\Im{[\tilde{C}(\textbf{r})+S(\textbf{r})]}\right)^2,
\end{eqnarray}
where $\Re$ and $\Im$ denote the real and imaginary parts. Using the properties of circular Gaussian distributions, $\Re[\tilde{C}(\textbf{r})+S(\textbf{r})]$ and $\Im[\tilde{C}(\textbf{r})+S(\textbf{r})]$ are independent Gaussian random variables of same variance $I_s/2$.
We can rewrite the intensity with real and imaginary terms of variance unity:
\begin{equation}
I =\frac{I_s}{2}\left(\left(\Re{[\sqrt{2I_s^{-1}}\tilde{C}(\textbf{r})+S(\textbf{r})]}\right)^2+
\left(\Im{[\sqrt{2I_s^{-1}}\tilde{C}(\textbf{r})+S(\textbf{r})]}\right)^2\right) = \frac{I_s}{2} \tilde{I},
\end{equation}
where $\var[\Re[\sqrt{2I_s^{-1}}\tilde{C}(\textbf{r})+S(\textbf{r})]]=\var[\Im[\sqrt{2I_s^{-1}}\tilde{C}(\textbf{r})+S(\textbf{r})]]=1$.\\
The random variable $\tilde{I}$ follows a de-centered $\chi^2$ with two degrees of freedom: $\chi^2_{2}(m)$, with a decentering parameter $m=2I_s^{-1}I_c$, \cite[chap. 29]{JKB95}.
The probability density function for $\tilde{I}$ is therefore:
\begin{equation}\label{Eq:hypergeom1}
\mathcal{P}(v)=2^{-1} e^{-(m+v)/2} f_{1}\left( \frac{1}{4}m v\right), \; v>0,
\end{equation}
where $f_q(z)$ is the regularized confluent hypergeometric function and $_0F_1(;q;z)$ the confluent hypergeometric function defined as:
\begin{equation}
f_q(z) =\sum_{n=0}^\infty \frac{1}{\Gamma(q+n)n!}z^n = \frac{\ _0F_1(;q;z)}{\Gamma(q)}
\end{equation}
Finally, the probability density function of the intensity $I = {I_s}/{2} \tilde{I}$ is:
\begin{equation}\label{EqRicianHyperGeom}
p_I(I)=\frac{e^{-\frac{I_c+I}{I_s}}}{I_s} \, _0F_1\left(;1;\frac{I_c \, I}{I_s^2}\right)
\end{equation}
This expression is equivalent \footnote{The Mathematica software \citep{W99} can be used to derive these expressions, and the equivalence between Eq.\ref{EqRicianHyperGeom} and Eq.\ref{EqRician} can be verified easily using the functions Simplify and FunctionExpand.} to the \textit{modified Rician distribution} derived by \citet{Goo75} and used by \citet{CC98,CC99,CC00,CC01}:
\begin{equation}\label{EqRician}
p_I(I)=\frac{1}{I_s}\exp\left(-\frac{I+I_c}{I_s}\right)\mathrm{I_0}\left(\frac{2 \sqrt{I} \sqrt{I_c}}{I_s}\right),
\end{equation}
where $\mathrm{I_0}$ denotes the zero-order modified Bessel function of the first kind.
The Rician distribution is illustrated in Fig.\ref{figrician}. A comparison between the Rician model and simulation data will be presented in Sec.\ref{TestRicianSims}.

An interesting particular case is when the background $\tilde{C}(\textbf{r})$ is zero and only the speckle term is present. Making $I_c=0$ in Eq.\ref{EqRician} (this happens at the zeros of the perfect PSF or using a perfect coronagraph), the PDF reduces to:
\begin{equation}\label{EqExpDistrib}
p_I(I)=\frac{1}{I_s}\exp\left(-\frac{I}{I_s}\right).
\end{equation}
This PDF corresponds to the well-known negative exponential density for a fully developed speckle pattern (e.g. laser speckle pattern) \citep{Goo00}. 
Finally, the distribution at photon counting levels can be obtained performing a Poisson-Mandel transformation of the high flux PDF in Eq.\ref{EqRician}. An analytical expression of this PDF has been given in \citep{AS04a}.


The mean and variance of the intensity can be obtained by several ways. A first method \citep{Goo75,Goo00}, is to express the mean intensity $\esp[I]$ and the second order moment of the intensity $\esp[I^2]$ as a function of $C(\textbf{r})$ and $S(\textbf{r})$. The second order moment for the intensity is the fourth order moment for the complex amplitude: $\esp[ I^2] =\esp [(C+S)(C^{*}+S^{*})]^2]$ (omitting the variables $\textbf{r}$ for clarity), which can be simplified using the properties of Gaussian distributions, with $\esp[ S S^{*} S S^{*}]= 2 \esp[ S S^{*}] \esp[ S S^{*} ]=2 I_s^2$ we obtain: $\esp[I^2]=I_c^2+4 I_c I_s+2 I_s^2$. 
A second method is to derive a general analytical expression for the moments of the Rician distribution. This can be be obtained either from the definition of the moments of Eq.\ref{EqRician} \citep{Goo75}, or computing the derivatives of the moment generating function \citep{AS04a}.
The instantaneous intensity in the focal plane (Eq.\ref{Eq:InstantaneousInt}) can be written as:
\begin{equation}\label{EqIntensInst}
I=|C(\textbf{r})|^2+|S(\textbf{r})|^2+2 \mathrm{Re}[C^*(\textbf{r})\,S(\textbf{r})].
\end{equation}
Since $\esp[S(\textbf{r})^*]=\esp[S(\textbf{r})]^*=0$ (circular Gaussian distribution), the mean intensity is simply the sum of the deterministic diffraction pattern with a halo produced by the average of the speckles: $I_c+I_s$ or $\tilde{I_c}+I_s$, respectively for direct and coronagraphic images. The variance also finds a simple analytical expression, and we have:
\begin{eqnarray}\label{Eq:moyvar}
\esp[I]&=&I_s+I_c\nonumber\\
\sigma^2_I&=&I^2_s+2 I_s I_c.
\end{eqnarray}
The variance associated with photodetection can be added to this expression to obtain the total variance $\sigma^2=\sigma^2_I+\sigma^2_P$,
where $\sigma^2_P$ is the variance associated to the poisson statistics: $\sigma^2_P=I_c+I_s$. \\
The total variance is therefore:
\begin{equation}\label{Eq:VarSpeckPhot}
\sigma^2=I^2_s+2 I_s I_c+I_c+I_s.
\end{equation}

\begin{figure}[htbp]
\center
\resizebox{.6\hsize}{!}{\includegraphics{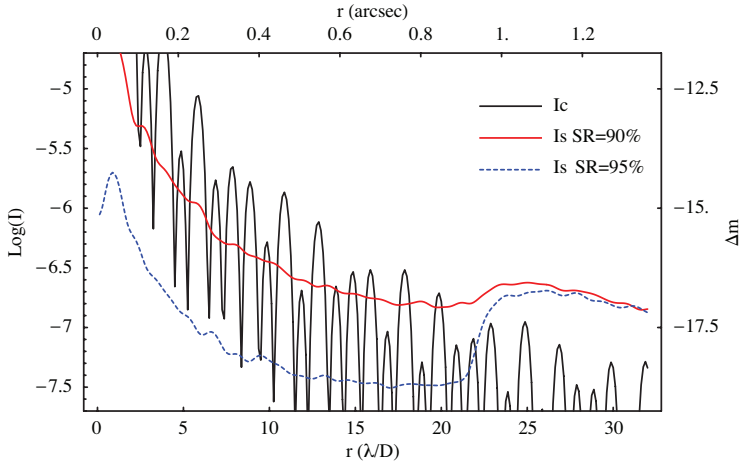}}
 \caption{Numerical simulation to illustrate the decomposition of the mean intensity PSF into two components $I_{c}$ and $I_s$ for two Strehl Ratios $90\%$ (V=8) and $95\%$ (V=4) for a direct, non-coronagraphic image. The $I_s$ term corresponds to the mean speckle halo and the $I_{c}$ term corresponds to the perfect PSF, scaled to the Strehl Ratio, so that the total intensity remains normalized (the difference between the two $I_c$ profiles is neglected here in log scale). The simulation is made with PAOLA.}
 \label{figIcIs}
\end{figure}

\begin{figure}[htbp]
\center
\resizebox{.6\hsize}{!}{\includegraphics{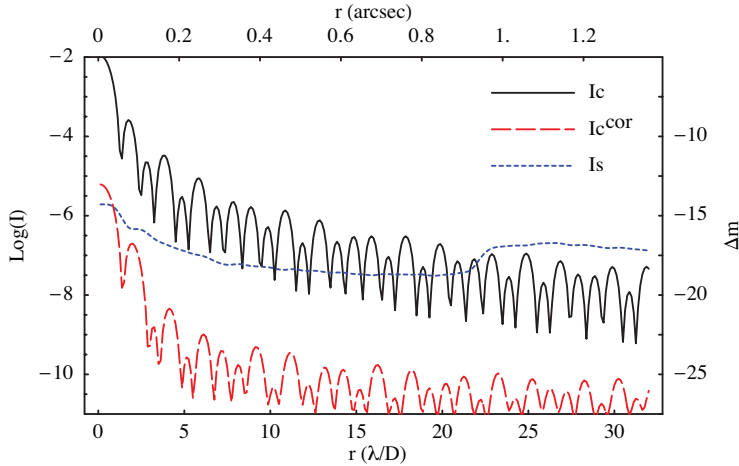}}
 \caption{Effect of a coronagraph on the $I_{c}$ term while the $I_s$ term is assumed unmodified (Sec.\ref{Sec:Prop}). We only consider one case of AO in this figure (SR=95\%) and illustrate the $I_c$ term for the direct and coronagraphic case. The effect of the APLC coronagraph here is to reduce the perfect PSF below the speckle halo. The corresponding long exposure image is totally dominated by the halo and no residual ringing remains.}
 \label{figIcIscorono}
\end{figure}

In the case of direct images, the term $I_c$ corresponds to the perfect Point Spread Function (PSF) scaled to the SR. 
In the case of coronagraphic images, the focal plane intensity is not invariant by translation, and therefore it is technically not a true PSF. However, we will use the term ``coronagraphic PSF'' for simplicity and to follow the general usage in the community.
The term $I_s=\esp[|S(\textbf{r})|^2]$ is a function of the radial distance $r$, which can describe an actual AO halo. These PSFs and halo structures have been studied analytically \citep{Mof69,Rac96,RWN99}. It is also possible to determine the halo profile directly from numerical simulations, and an illustration of $I_c$ and $I_s$ is shown in Fig.\ref{figIcIs}. The long exposure PSF profile is the sum of these two contributions, the halo clearing effect for higher SR \citep{SKM01} is clearly visible between the two figures. The shape of the halo is due to the spatially filtered wavefront sensor \citep{PM04} used in this simulation.
In Fig.\ref{figIcIscorono} we show the effect of a coronagraph on the $I_{c}$ term, while the $I_s$ term is left unmodified as explained in Sec.\ref{Sec:Prop}. In this figure we only consider one of the previous two AO cases. In this example, the coronagraph is good enough to render the constant background term $I_{c}$ negligible when compared to the speckle term $I_s$.

\subsubsection{Effect of a coronagraph on speckle pinning}

In a direct, non coronagraphic image, the term coupling the deterministic $C(\textbf{r}) $ and random parts $S(\textbf{r}) $ in Eq.\ref{EqIntensInst} corresponds to the so-called ``speckle pinning'', discussed by several authors  \citep{BDT01,SLH02,Blo03,SHM03,PSM03,Blo04}, using an expansion of $|\mathcal{F}[P(\textbf{r}) e^{j \varphi(\textbf{r})}]|^2$. In this expansion approach, the PSF consist of a sum of terms, some of which contain a multiplicative factor $\mathcal{F}[P(\textbf{r})]$, corresponding to the diffracted field (the Airy amplitude for a circular aperture). These terms inherit the zeros of $\mathcal{F}[P(\textbf{r})]$ and contribute all together to the pinned speckles. Pinned speckles therefore have zero amplitude at the Airy nulls, and non-pinned speckles remain at these locations. The first order PSF expansion term, denoted by $p_1$ in \citet{PSM03}, corresponds to the pinned speckles and the second order ($p_{2 halo}$) to non-pinned speckles. Higher order terms contribute to pinned and non-pinned speckles. This expansion approach provides particularly interesting insight into the spatial properties and symmetries of the speckles for each order of the expansion \citep{SHM03,PSM03,Blo04}.
The expansion and our decomposition are therefore very similar. In Eq.\ref{EqIntensInst}, our pinning term $2 \mathrm{Re}[C^*(\textbf{r})\,S(\textbf{r})]$ includes the contribution of all pinning terms from the infinite expansion. However, our constant term in the decomposition is not 1 and this term can include the effect of static aberrations, as discussed below.

The analysis of the statistical properties of the speckles enables a deeper understanding of the pinning phenomenon.
As shown in Eq.\ref{EqIntensInst} and Eq.\ref{Eq:moyvar}, the pinned speckle term of Eq.\ref{EqIntensInst} does not contribute to the mean intensity, but only contributes to the variance.
A numerical simulation is used in Fig.\ref{figrician} to illustrate the Rician distribution for a direct image, at three different positions in the field: one at the top of an Airy ring (strong pinning effect), one at a PSF zero (no speckle pinning) and one at an intermediate position. 
Speckle pinning can be well illustrated by the analysis of these PDFs, as speckle intensity and fluctuations are amplified by the term $I_c$. This can be seen in the PDFs in Fig.\ref{figrician}, where the widths increase with $I_c$.
Depending on the amplitude of the Airy pattern at successive rings, the intensity $I_c$ is alternatively large and small and the variance of the speckles is amplified accordingly by the coherent part of the wave amplitude $C(\textbf{r})$, with corresponding intensity $I_c$. At the zeroes of the PSF, no amplification occurs and the statistics is equivalent to that of a fully developed speckle pattern (exponential statistics). It is important to note that speckles fluctuations are not fully cancelled there (Fig.\ref{figrician}), but simply \textit{not amplified} and their statistics is that of laser speckles. 

\begin{figure}[htbp]
\center
\resizebox{.6\hsize}{!}{\includegraphics{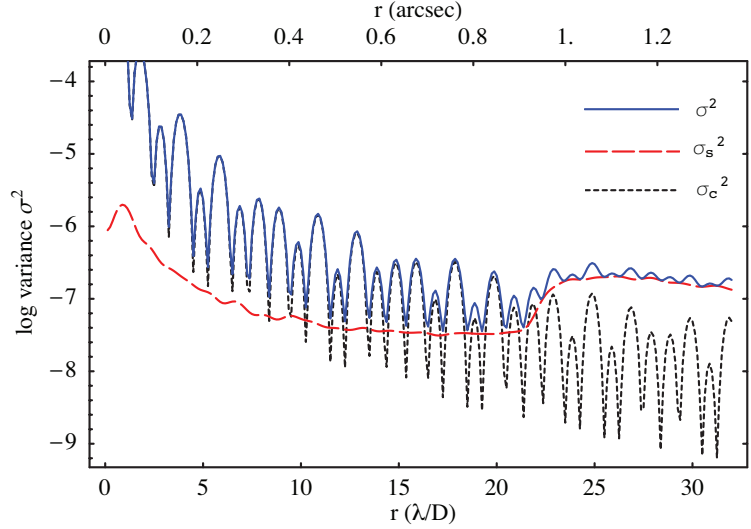}}
 \caption{Total variance for direct imaging (blue solid line) with contributions of the variance terms $\sigma_s$ (red dashed line) and $\sigma_c$ which correspond to the pinning contribution (black dotted line). In direct imaging, the variance budget is totally dominated by the speckle pinning effect, at least close to the axis. This simulation is made for a $SR=95\%$. }
 \label{Fig:sigmaCsigmaS}
\end{figure}

\begin{figure}[htbp]
\center
\resizebox{.6\hsize}{!}{\includegraphics{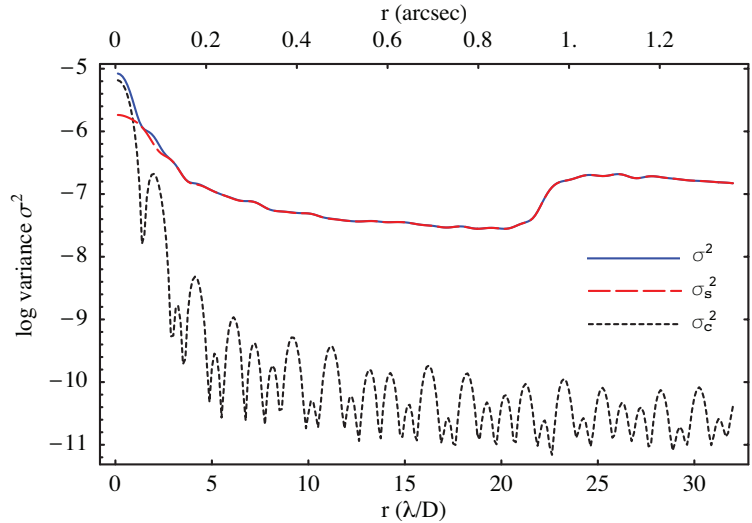}}
 \caption{Total variance for coronagraphic imaging (blue solid line), with contributions of the variance terms $\sigma_s$ (red dashed line) and $\sigma_c$ which correspond to the pinning contribution (black dotted line). A coronagraph can remove the speckle pinning contribution to the variance of the noise. This simulation is made for a $SR=95\%$ and using and Apodized Pupil Lyot Coronagraph.}
 \label{Fig:sigmaCsigmaScoro}
\end{figure}

The effect of a coronagraph on speckles can be well explained from this statistical modeling.
Formally, as shown in Eq.\ref{psi4}, the effect of a coronagraph is to replace the telescope PSF $I_c$ by the on-axis coronagraphic PSF after a coronagraph $\tilde{I_c}$. This term includes the effect of static aberrations if they are included in the model.
\begin{itemize}
\item In the perfect case of an ideal coronagraph achieving a total extinction of the star \citep{RRB00,ASF02,KT02,FPS05}, speckle pinning is fully canceled if there are no static aberrations in the system. 
\item In the case of static aberrations in the pupil (due to polishing and alignment errors for example), these aberrations propagate through the coronagraph, according to the description given in Sec.\ref{Sec:Prop}, or in \citet{SSO07}. The deterministic response of the coronagraph $\tilde{I_c}$ (Eq.\ref{Eq7}) therefore includes the effect of these static aberrations. As a result, static aberrations leaking through a coronagraph contribute to speckle pinning, \emph{even in the case of a perfect coronagraph}. Such effect can be produced for example by dead actuators on the deformable mirror \citep{SOP06}. 
\end{itemize}
Following \citet{AS04}, we can illustrate the effect of a coronagraph on speckle noise, by breaking down the total variance (Eq.\ref{Eq:VarSpeckPhot}) into two contributions, $\sigma_c^2=2 I_c I_s+I_c$ and $\sigma_s^2=I_s^2+I_s$. The term $\sigma_c^2$ is the part of the variance that can be removed by a coronagraph by changing $I_c$, thus affecting speckle pinning. In Fig.\ref{Fig:sigmaCsigmaS} and Fig.\ref{Fig:sigmaCsigmaScoro} we illustrate these variance contributions for direct and coronagraphic images. In the case of the direct image, the ringing aspect of the variance corresponds to speckle pinning (where the variance is amplified). The coronagraphic variance profile is smooth (no amplification of the speckles is seen at the location of the diffraction rings).
Details of the numerical simulation are given in Sec.\ref{Sec:NumSims}.

\subsubsection{Test of the Rician distribution with numerical simulations}\label{TestRicianSims}

Tests of the Rician statistics on real data using the Lick observatory AO system have been carried out by \citet{FG06}, showing that the Rician model is consistent with the data.
Several complications exist in real data, so we test the Rician distribution on numerical simulations to determine whether the model is acceptable in a simple case consistent with the hypothesis used in the physical model, without any additional complicating circumstances or noise.
We used PAOLA to generate 10000 AO-corrected instantaneous phase screens corresponding to an ExAO system, on a 8m telescope (we used a telescope geometry compatible with Gemini or VLT). The parameters chosen for this simulation include 44 actuators across the pupil, an integration time and time lag of 0.5ms for a magnitude $V=8$ star, and observations in the H-band. The atmosphere include a typical $C_n^2$ profile for Cerro-Pachon and the seeing is assumed to be 1.4 arcec. The Strehl Ratio of these simulated images is $83\%$. 

\begin{figure}[htbp]
\center
\resizebox{.5\hsize}{!}{\includegraphics{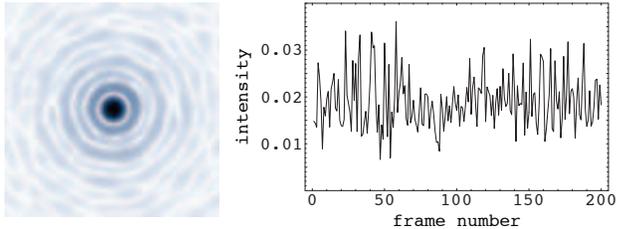}}
 \caption{Example of instantaneous PSF simulated with PAOLA, including both phase and amplitude errors (left) with the intensities for 200 independent frames at an arbitrary position in the focal plane (right). 10000 independent PSFs have been generated at a sampling of 3 times Nyquist, and the values along a radial axis are stored for this study.}
 \label{FigPSFsims}
\end{figure}

In each image, we extract the intensities values along a radius to construct 50 intensity series in the focal plane at these 50 pixel locations. An example of the first 200 intensity values at an arbitrary location is given in Fig.\ref{FigPSFsims}.

\begin{figure}[htbp]
\center
\resizebox{.7\hsize}{!}{\includegraphics{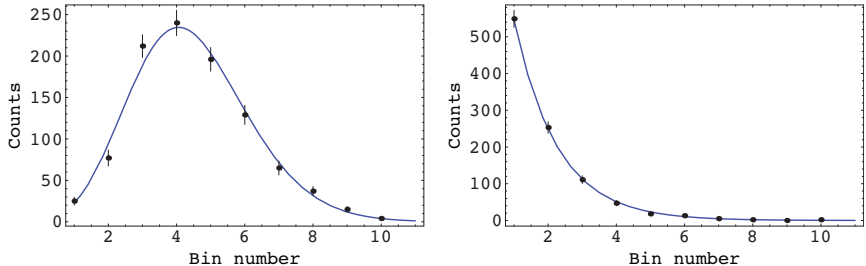}}
 \caption{Examples of fitted Rician distributions to the binned data, in two cases corresponding to two different locations in the focal plane: At a top of an Airy ring (left) and at a zero of the perfect PSF where the distribution is exponential (right). 1000 independent realizations have been used in this simulation.}
 \label{FigFitExample}
\end{figure}

For each of these 50 points, we performed a Maximum Likelihood estimation of the parameters $I_c$ and $I_s$, assuming a Rician distribution for the data. The Likelihood has been computed as function of the two parameters for the un-binned data and maximized using optimization routines of Mathematica.
We then perform the $\chi^2$ and Kolomogorov-Smirnoff test statistics on these results. We use ten identical identical bins for the $\chi^2$ test. In Fig.\ref{FigFitExample}, we show two examples of binned data with error bars (here due to the Poisson statistics), superimposed with the Rician distribution for the estimated $I_s$ and $I_c$ parameters.

Our implementation of the Kolmogorov-Smirnoff test is based on a Monte-Carlo estimation of the distribution. Indeed, when parameters are estimated from the data (here $I_c$ and $I_s$), the distribution of the KS test is not known analytically and must be estimated from Monte-Carlo simulations. For that, we calculated a distribution of KS test values for 100,000 random drawings of random Rician data where the parameters have to be estimated from the data. This empirical distribution of the KS test values is used to generate the empirical right tail values for the test statistic.

\begin{figure}[htbp]
\center
\resizebox{.7\hsize}{!}{\includegraphics{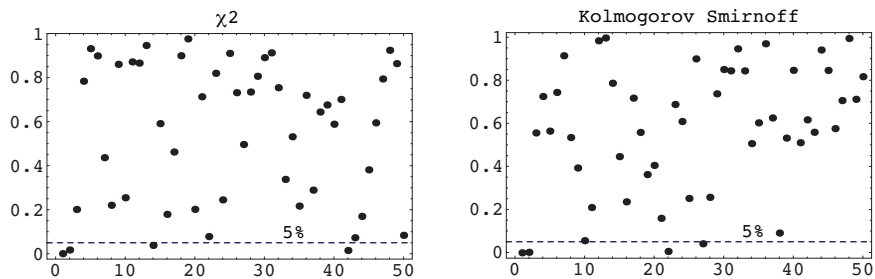}}
 \caption{$\chi^2$ and KS test results for the 50 locations along a radius in the focal plane. The first point corresponds to the on-axis case, and the last point is located at $8.3\lambda/D$, with a separation of $\lambda/(6D)$ between each point. The solid line shows the $5\%$ rejection threshold. The first two points are rejected by both tests, which illustrates the non-Rician distribution. In the rest of the field both tests are consistent with the Rician distribution hypothesis at the $5\%$ rejection level. }
 \label{FigKSchi2results}
\end{figure}

The results of the two tests (right tail values) are given in Fig.\ref{FigKSchi2results}, for each of the 50 points along the radial axis (between 0 and $8.3\lambda/D$). It is interesting to note that for the first two points (on-axis intensity distribution), both tests conclude that the Rician model is incompatible with the data at the 5\% level. Although the statistical properties at the central point are not directly relevant for evaluating the detection limits in high contrast images, this question is important in itself as it corresponds to the distribution of Strehl Ratio. This problem has been studied by \citet{SF07} where it is shown that the distribution at the center of the image is given by a ``reversed'' non-central Gamma distribution. This problem was studied independently by \citet{GCR06,G06,CGR06} with similar results.

\begin{figure}[htbp]
\center
\resizebox{.5\hsize}{!}{\includegraphics{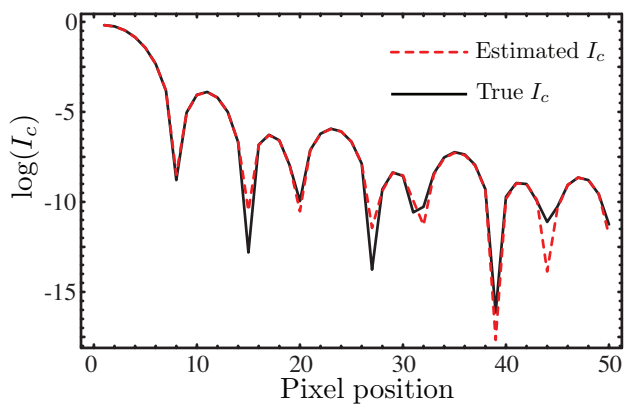}}
 \caption{Comparison between the actual perfect term $I_c$ from the simulation without wavefront errors (red dashed lines), and the $I_c$ values retrieved from the Maximum Likelihood estimation of the parameters (solid line), assuming the Rician distribution for the intensity. Both curves show a very good agreement, consistent with the test statistics results. Log intensities are plotted as a function of the radial position in pixels for easier comparison with other figures.}
 \label{FigIcFromStat}
\end{figure}

Outside of the center, most pixels pass the two tests except for about 2 pixel locations, which are rejected, by one test or the other. This result is consistent with the 5\% level we chose, given the total number of locations (50 pixels) and we conclude that the Rician model is compatible with the simulated data. We have verified on a few sets of simulations that the location of these pixel failing the test is not relevant and simply due to the statistics. On the contrary, the central pixel fails the tests systematically.
It is interesting to compare the estimated parameters with the actual parameters known from the simulation. In Fig.\ref{FigIcFromStat} we compare the actual $I_c$ profile known from the simulation data and compare it to the reconstructed $I_c$ profile from the estimation of the Rician statistics. 
Both curves show a very good agreement, even somewhat surprisingly at the center where the Rician model is wrong. 
This result is compatible with the satisfactory results of the test statistics. \citet{FG06} discussed this type of approach as a possible way to reconstruct a telescope perfect PSF from the statistics of the PSFs. They note that this procedure might be difficult to implement with real data, but is at least verified in simulations in this simple case. 

\subsection{Effect of quasi-static aberrations on the statistical properties}\label{SecQuasiStatic}

Observations at high contrast have shown that the main dynamic range limitations are due to long-lived quasi-static speckles \citep{BML97,OGK01,MDN03,BCL03,BRB04,SOH06}. Practical solutions have been proposed to overcome the quasi-static speckles and will be implemented on the next generation of high contrast imagers (GPI, Sphere) \citep{KBR06,MLD06} or considered for space projects \citep{BT06,GKV06}. In ExAO coronagraphic images, \citet{HOS06} have studied residual speckle lifetimes, and evidenced two types of speckles, one with lifetimes of a few seconds and the other with lifetimes of a few hundred seconds. Although there is no reason to believe the universality of these particular values, this problem of quasi-static speckles is general enough to need further understanding and theoretical insight. Quasi-static speckles are produced by the slow deformations of the optics (thermal and mechanical) that happen for example as the telescope slews to follow the star. The typical time scale for these slow variations of aberrations are tens of minutes. 

We can generalize the approach of Sec.\ref{Sec:Prop} to include two types of aberrations with different time scales. We will assume that the error wavefront at the entrance pupil consists of the coherent addition of two terms: a fast-varying wavefront related to atmospheric residuals, and a quasi-static aberration term. We assume that quasi-static aberrations can be described as the sum of a deterministic static aberration and a slowly evolving, zero-mean complex aberration: the static aberrations correspond to the polishing errors and aberrations produced by the optical design and actual misalignments in the system. Quasi-static aberrations corresponds to slow zero-mean fluctuations around this static (and therefore deterministic) contribution.
Under these hypothesis, we can add a fourth term to the wavefront decomposition of Eq.\ref{Eq1}:
\begin{equation}
\Psi_{1}(\textbf{r})=A+A_s(\textbf{r})+a_{1}(\textbf{r})+a_{2}(\textbf{r}),
\end{equation}
where, as previously, A corresponds to the perfect wave and $A_s(\textbf{r})$ to the static aberrations (deterministic). The error terms $a_{1}(\textbf{r})$ and $a_{2}(\textbf{r})$ correspond respectively to the AO-corected and quasi-static aberrations. 
We assume that the lifetime of $a_{1}(\textbf{r})$ and $a_{2}(\textbf{r})$ are respectively $\tau_1$ and $\tau_2>\tau_1$.
We consider the longest lifetime $\tau_2$ as the time unit, and we denote by $N$ the number of fast-speckle realizations during a slow-speckle lifetime:
\begin{equation}
N=\frac{\tau_2}{\tau_1}=\frac{N_1}{N_2},
\end{equation}
where $N_{1}$ and $N_{2}$ are the number of realizations for each processes during a long exposure of duration $\Delta t$.
Following the results of the previous sections, we can write the field amplitude in the focal plane during $N$ successive short time intervals $\tau_1$ as:
\begin{equation}
\Psi_{4}(\textbf{r})=\tilde{C}(\textbf{r})+S_1^{[k]}(\textbf{r})+S_{2}(\textbf{r}),\; k=1, \ldots, N.
\end{equation}
As previously, $\tilde{C}(\textbf{r})=C(\textbf{r})+S_s(\textbf{r})$ where $C(\textbf{r})$ corresponds to the focal plane field (direct of coronagraphic) in the perfect case without aberrations, and $S_s(\textbf{r})$ to the static aberration (e.g. polishing errors or actual optical design aberrations).
$S_1^{[k]}(\textbf{r})$ corresponds to the AO-corrected atmospheric aberrations and $S_{2}(\textbf{r})$ to the quasi-static aberrations.
Outside of the central point in the focal plane, $S_1^{[k]}(\textbf{r})$ and $S_{2}(\textbf{r})$ are independent circular Gaussian distributions with zero mean: $S_1 ^{[k]}(\textbf{r})\sim \mathcal{N}_c(0,I_{s1})$ and $S_2(\textbf{r}) \sim \mathcal{N}_c(0,I_{s2})$ with respective lifetimes $\tau_{1}$ and $\tau_{2}$. Also, the spatial properties of $S_{1}^{[k]}(\textbf{r})$ and $S_{2}(\textbf{r})$ are different as they result from different aberrations sources, and their corresponding variances $I_{s1}$ and $I_{s2}$ can be generated considering the appropriate spatial power spectra.

During a time unit $\tau_2$ corresponding to the lifetime of a slow speckle, the intensity is:
\begin{eqnarray}\label{Eq:IntensQS}
I &=& \sum_{k=1}^{N}| \tilde{C}(\textbf{r})+ S_1^{[k]}(\textbf{r}) + S_2(\textbf{r}) |^2\nonumber\\
&=& \sum_{k=1}^{N}| U_k|^2.
\end{eqnarray}

The expected value of $| U_k|^2$ is $\esp[|U_k|^2] = \var[U_k]+|\esp[U_k]|^2$,
where $U_k$ follows the Gaussian distribution $U_k \sim \mathcal{N}_c(\tilde{C}(\textbf{r}),I_{s2}+I_{s1})$.
Therefore, the mean intensity is:
\begin{equation}\label{Eq:EspIntQS}
\esp[I] = N (I_{s2}+I_{s1}+I_c).
\end{equation}
Reminding that $I_c=| \tilde{C}(\textbf{r})|^2$. This expression is consistent with the simple case with one type of speckle (Eq.\ref{Eq:moyvar}), as it is expressed here for an exposure $\tau_2=N \tau_1$.

The variance of the intensity is defined as:
\begin{eqnarray}\label{Eq:VarIntensQS}
\var[I] &=& \sum_{k=1}^{N} \var[|U_k|^2]+ \sum_{k\not = l} \mathsf{cov}[|U_k|^2,|U_l|^2]\nonumber\\
&=&N\var[|U_k|^2]+N(N-1)\mathsf{cov}[|U_k|^2,|U_l|^2].
\end{eqnarray}
The covariance $\mathsf{cov}[|U_k|^2,|U_l|^2]$ can be easily computed using the expansion of the high order moments of a complex Gaussian vector as a function of its 
high order cumulants \citep{MC87,F06}.
The covariance finds a simple expression:
\begin{equation}
\cov[|U_k|^2,|U_l|^2] = |\cov[U_k,U_l^\ast]|^2+
\esp[U_k^\ast]\esp[U_l]\cov[U_k,U_l^\ast]+ 
\esp[U_k]\esp[U_l^\ast] \cov[U_l,U_k^\ast].
\end{equation}
With $U_k \sim \mathcal{N}_c(\tilde{C}(\textbf{r}),I_{s2}+I_{s1})$ and $k\not = l$
$\cov[U_k,U_l^\ast] = I_{s2}$, we obtain immediately the covariance and the variance (for k=l):
\begin{eqnarray}
\cov[|U_k|^2,|U_l|^2] = I_{s2}^2+2 I_c I_{s2}\nonumber\\
\var[|U_k|^2] = (I_{s2}+I_{s1})^2 +2 I_c (I_{s2}+I_{s1})
\end{eqnarray}
Finally, the variance of the intensity (Eq.\ref{Eq:VarIntensQS}), for an exposure $\tau_2$, becomes:
\begin{eqnarray}\label{Eq:VarQS1}
\var[I]&=&N(I_{s1}^2+N I_{s2}^2 + 2 I_c(I_{s1}+N I_{s2})+ 2 I_{s1} I_{s2})\nonumber\\
	 &=&N\,\sigma_I^2,
\end{eqnarray}
where the variance $\sigma_I^2$ for a short exposure $\tau_1$:
\begin{equation}\label{Eq:sigmaI}
\sigma_I^2=I_{s1}^2+N I_{s2}^2 + 2 I_c(I_{s1}+N I_{s2})+ 2 I_{s1} I_{s2}
\end{equation}
is consistent with the case with one type of speckle (Eq.\ref{Eq:moyvar}), if $I_{s2}=0$. 
\begin{itemize}
\item
In this generalized expression of the variance, the pinning term $2 I_c(I_{s1}+N I_{s2})$, still exist.
However, it now includes a term corresponding to the pinning of quasi-static speckle. 
The coefficient $N$ (the ratio of the speckle lifetimes) can be very large, which makes this pinning particularly efficient: a quasi-static halo $I_{s2}$ which is $N$ times lower than $I_{s1}$ produces the same pinning effect in direct images.
This pinning contribution is still directly affected by a coronagraph, which reduces (or cancels) the $I_c$ term. 
In the case of a real, imperfect coronagraph, although no atmospheric pinning might be present, residual pinning of quasi-static speckles may occur due to the amplification by the factor $N$ if the quasi-static aberrations are not sufficiently small.
\item
The cross term $2 I_{s1} I_{s2}$ corresponds to speckle pinning of the atmospheric by quasi-static aberrations (coherent amplification of the atmospheric speckles by the quasi-static speckles).
If the system is dominated by the quasi-static speckles ($I_{s2}$ large), this term is negligible compared to the halo $N I_{s2}^2 $.
\end{itemize}
It is possible to specify the performance of a coronagraph to avoid speckle pinning, using this expression for the variance. Pinning terms are negligible compared to the halo terms if: 
\begin{equation}
2 I_c(I_{s1}+N I_{s2})\ll I_{s1}^2+N I_{s2}^2 ,
\end{equation}
assuming the cross term negligible.
In the case of only one type of speckles, or when the system is dominated by quasi-static speckles, this condition simply becomes $I_c\ll I_s$ \citep{AS04}.
The necessary rejection of the coronagraph can be defined as a ratio of the coronagraphic PSF $I_c^{coro}$ to the direct PSF $I_c^{direct}$.  A similar approach was used by \citet{Blo04b} to quantify the coronagraph effect on pinned speckles using the expansion approach. 

\section{Dynamic range in coronagraphic images}\label{DynRangeSec}

In this paper we adopt a definition of the dynamic range based on an expression of the Signal to Noise Ratio (S/R), assuming a typical $5\sigma$ level detection limit. 
Based on the knowledge of the noise statistics, more advanced methods are possible, where the dynamic range can be defined in terms of probability of detection and false alarm \citep{MF03,FCS06}. In particular, because of the non-Gaussian statistics (exponential distribution for a perfect coronagraph), confidence levels have to be carefully defined, as they do not correspond to the usual values for a Gaussian distribution, C. Marois et al. (2007, in preparation).
In this section we define analytically the dynamic range in a coronagraphic experiment, using the variance expressions obtained above. This approach enables a semi-analytical method. We present some simulations results for the dynamic range. 

	\subsection{Signal to Noise ratio and dynamic range}

For simplicity in the calculation of the variance (Sec.\ref{SecQuasiStatic}), we expressed the intensity for a time unit $\tau_2$ as a sum of short exposures $\tau_1$ (Eq.\ref{Eq:IntensQS}). In the expression of the variance we derived (Eq.\ref{Eq:VarQS1}), the terms $I_{s1}$, $I_{s2}$, $I_{c}$ correspond implicitly to a number of photons during an exposure $\tau_1$. However, it is convenient in practice to define normalized intensity terms for a single photon at the entrance aperture and scale them using the star flux $F_\star$. A normalization term $\tau_1^2 \,F_\star$ is therefore necessary in order to use such normalized intensities. Recalling that a long exposure consists of $N_2$ exposures of durations $\tau_2$, we obtain the final expression of the variance for a long exposure $\Delta t$:
\begin{eqnarray}\label{Eq:VarQS2}
\sigma_{speckle}^2&=&N_2\, \tau_1^2 \, F_\star^2 \,\var[I] \nonumber\\
			 &=&N_1 \, \tau_1^2 \, F_\star^2 \,(I_{s1}^2+N I_{s2}^2 + 2 I_c(I_{s1}+N I_{s2})+ 2 I_{s1} I_{s2}).
\end{eqnarray}
The photon noise is obtained by applying the same normalization, multiplying Eq.\ref{Eq:EspIntQS} by $ N_2 \tau_1^2$:
\begin{equation}\label{Eq:VarQS3}
\sigma ^2_{photon}=\left(I_c+I_{s 1}+ I_{s 2}\right)\,F_\star\, \Delta t, 
\end{equation}
and introducing $\sigma_p^2=I_c+I_{s 1}+ I_{s 2}$ for consistency with the case with one type of speckle.
The signal from a companion is $F_p \, f(\theta) \, \Delta t$, where $F_p $ is the planet flux and $f(\theta)$ is a function that describes the off-axis response of the coronagraph to a companion. $f(\theta)$ can be computed independently. For simplicity, we use the maximum intensity of the off-axis companion for the signal, but other approaches such as matched filtering can also be used \citep{SOH06}. The signal to noise ratio (S/N) is:
\begin{equation}\label{Eq:SNR}
S/N = \frac{F_p \, f(\theta) \,\Delta t}
{\sqrt{N_1\, \tau_1^2 \,F_\star^2\, \sigma_I^2+ F_\star\, \Delta t\, \sigma_p^2}}.
\end{equation}
This expression for the signal to noise ratio can be immediately converted into an expression for the dynamic range, assuming a given S/N level. The dynamic range, denoted $d$ is simply the intensity ratio between an off-axis companion and the star that produces a given S/N level: $F_p=d\, F_\star$.
Since $\Delta t = N_{1}\tau_{1}$, we have:
 \begin{equation}\label{Eq:DR}
 d= \frac{S/N}{f(\theta)} \,\sqrt{\frac{\tau_1}{\Delta t}} \,\sqrt{\sigma_{I}^2+\frac{\sigma ^2_{p}}{\tau_1 F_\star}}.
 \end{equation}
Other noise contributions can be easily added in this expression if necessary, but a detailed study of any specific instrument is beyond the scope of this paper.


	\subsection{Semi-analytical method based on the statistical model}\label{Sec:NumSims}
The dynamic range (Eq.\ref{Eq:DR}), can be computed directly from numerical simulations of the terms $I_{s1}$, $I_{s2}$, and $I_{c}$.
We follow the equations used in the model (Eq.\ref{Eq:DefA} to Eq.\ref{psi4}) to construct numerical estimates for $I_c$ and $I_s$ (or $I_{s1}$, $I_{s2}$, if applicable).
The term $A$ is obtained from its definition (Eq.\ref{Eq:DefA}). We consider the average over the aperture and over a few independent realizations, generated with PAOLA \citep{JVC06} and including both amplitude $\rho(\textbf{r})$ and phase screens $\varphi(\textbf{r})$. As described in Sec.\ref{Sec:Prop}, the term $|A|^2$ corresponds to the Strehl Ratio, and is used to generate the $I_c$ profiles: direct and coronagraphic normalized PSFs are calculated in the perfect case without aberrations, and multiplied by $|A|^2$.

The term $S(\textbf{r})$ is generated following the notation of Eq.\ref{Eq6}: $S(\textbf{r})=\mathcal{F}[a(\textbf{r})\, P'(\textbf{r})]$, where $a(\textbf{r})=\rho(\textbf{r})\, \exp^{\imath\varphi(\textbf{r})}-A$ is the zero-mean complex amplitude at the aperture.
The $I_s$ term is obtained averaging independent realizations of $|S(\textbf{r})|^2$. It is easy to verify that the total intensity $I=I_c+I_s$ remains correctly normalized with this procedure.

Finally, radial profiles are generated by averaging these results azimuthally. Satisfactory profiles are obtained with only a few realizations (typically a few tens), where purely numerical simulations require large numbers of independent realizations to estimate the same variance. Once the various $I_c$ and $I_s$ terms have been generated, they can be readily combined according to Eq.\ref{Eq:VarQS2} or Eq.\ref{Eq:SNR} to produce variance or detection plots. Multiple instrumental cases can be easily studied this way, combining direct or various coronagraphic images with different cases of AO is done rapidly using our analytical expressions (for example to study the effect of stellar magnitude on the dynamic range). This method also enables the study of quasi-static speckles, which is otherwise computationally expensive. 

\begin{figure}[htbp]
\center
\resizebox{.6\hsize}{!}{\includegraphics{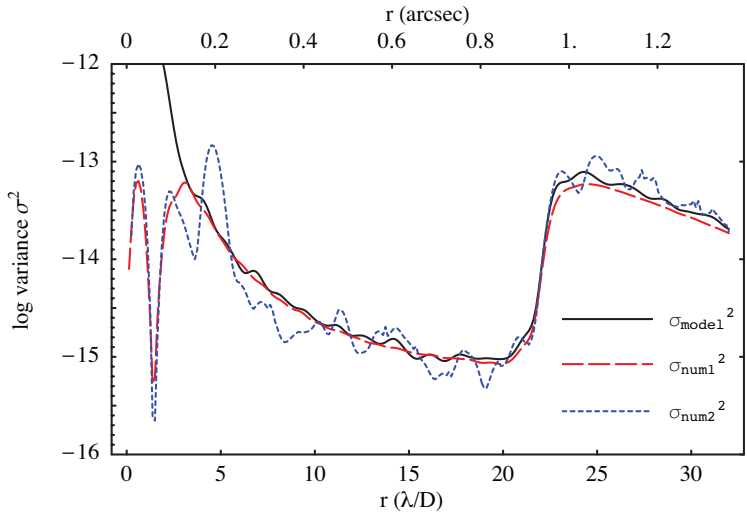}}
 \caption{Validation of the results obtained with the semi-analytical model, compared to a purely numerical experiment. The variance obtained from numerical estimation of $I_c$ and $I_s$ based on 20 realizations of phase and amplitude screens. Solid black line: model, dashed red line: variance for 1000 realizations, dotted blue line: variance estimated spatially on a single frame.}
 \label{Fig:CompVariance}
\end{figure}

We compared the variance obtained with the analytical model with the variance obtained from a purely numerical simulation, computing a large number $(\sim 1000)$ of independent realizations. This process is time-consuming because of the wavefront propagation through the coronagraph for each phase screen. Excellent agreement with a full numerical simulation of the variance is obtained with a few realizations $(\sim 20)$, using the semi-analytical method (Fig.\ref{Fig:CompVariance}). We note that the model gives an incorrect estimate of the variance behind the focal plane occulting mask. This is expected given the approximations described in Eq.\ref{Eq7} and Eq.\ref{psi4}. This is not a problem when studying dynamic range, as these expressions are weighted by the off-axis transmission $f(\theta)$ of the coronagraph, and results are only relevant outside the inner working angle of the instrument.

\subsection{Results and comparison with numerical simulations}

\begin{figure}[htbp]
\center
\resizebox{.6\hsize}{!}{\includegraphics{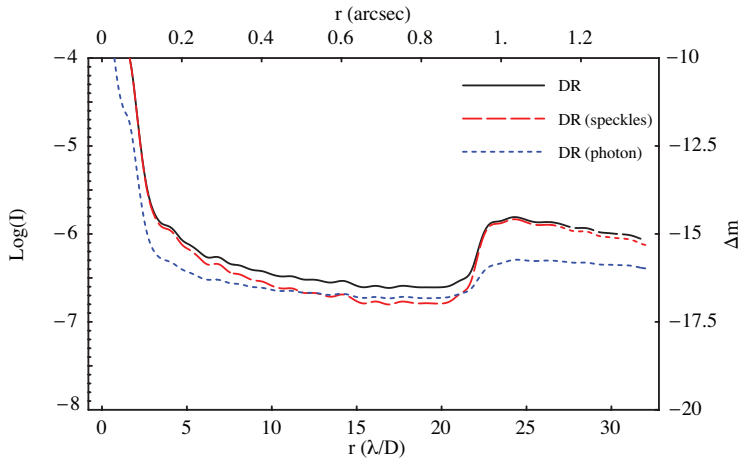}}
 \caption{Dynamic range simulation for a $1000s$ exposure (black solid line), showing the contributions of the photon noise alone (blue dotted line) and speckle noise alone (red dashed line). In this example, speckle and photon noise are balanced inside the control region and the outer region is dominated by the speckle noise.}
 \label{Fig:DRspecklephoton}
\end{figure}

We present some dynamic range simulation results, using the semi-analytical method based on Eq.\ref{Eq:DR}.
We consider an eight-meter telescope, an AO system with 44 actuators across the aperture, a spatially filtered wavefront sensor \citep{PM04}, and an open loop frequency of 2.5 kHz. The atmospheric simulation uses typical seeing values and $C_n^2$ profiles at Mauna Kea. 
A few cases have been simulated, with stellar magnitudes ranging from V=4 to V=8. Strehl Ratios are obtained between $87\%$ and $95\%$.
Atmospheric scintillation is also simulated as amplitude screens with PAOLA, which implements the method of \citet{R81}.
We assumed an instrumental throughput of $25\%$ and an arbitrary exposure time of $\Delta t=1000s$. The atmospheric lifetime is assumed to be $40ms$ for observations in the H-band and the stellar magnitude is V=4. The coronagraph in the simulation is an APLC similar to the one under study for GPI \citep{S05}.
With a $5\sigma$ detection level for this simulation, we illustrate the dynamic range (including photon and speckle noise only) in Fig.\ref{Fig:DRspecklephoton}. In this example, photon and speckle noise are approximately at the same level inside the AO control region. Outside the control region, the dynamic range is set by the speckle noise level.

It is interesting to note that the relative position of the speckle and photon contributions do not depend on exposure time (Eq.\ref{Eq:DR}). Indeed, only the speckle lifetime and stellar flux have an effect on the relative contributions of speckle and photon noise to the dynamic range. This is also true when both fast and quasi-static speckles are present. 
In terms of the contributions of $\sigma^2_s$ and $\sigma^2_c$ to the dynamic range as defined in \citet{AS04}, the APLC coronagraph is almost perfect in this simulation and the part of the variance corresponding to speckle pinning, $\sigma^2_c$, has been completely removed. The case illustrated in Fig.\ref{Fig:sigmaCsigmaS} and Fig.\ref{Fig:sigmaCsigmaScoro} corresponds to the same simulation and shows the effect of the APLC coronagraph canceling the pinning contribution completely.

Finally, we performed a simulation including both fast and quasi-static aberrations.
We considered for that a simple approach where the power spectrum of the quasi-static aberrations follows a power law $f^{-3}$, with a lifetime of 600s (10min). This choice is consistent with the assumption that quasi-static aberrations result from polishing errors, usually well described by such power laws. We considered both photon and speckle noise, according to Eq.\ref{Eq:DR}. In this example, although the quasi-static wavefront error is very small (10nm RMS), it dominates the error budget and limits the dynamic range inside the control radius of the AO system. This result depends directly on the power spectrum chosen to generate these aberrations. This could explain the results obtained with the Lyot project coronagraph where the dynamic range plots do not show the expected halo-clearing region within the control radius of the AO system \cite{HOS06}. A slightly shallower power spectrum, or a higher level of quasi-static aberrations would easily reproduce the observed dynamic range by filling the cleared region completely. This effect could be due mainly to the presence of broken actuators \citep{SOP06} in the AO system.
\begin{figure}[htbp]
\center
\resizebox{.6\hsize}{!}{\includegraphics{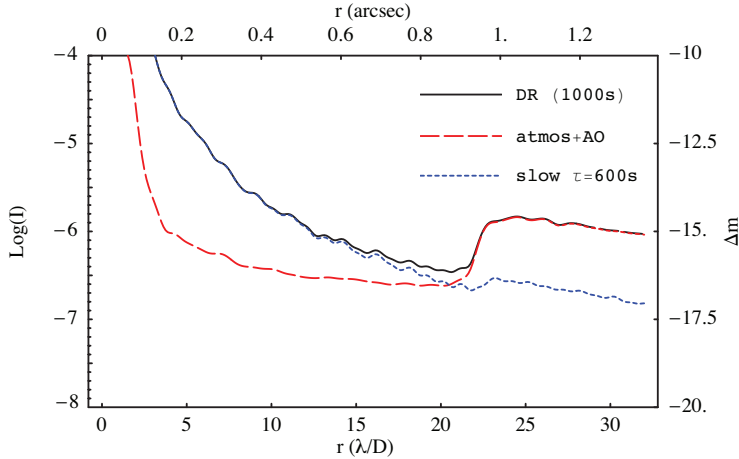}}
 \caption{Dynamic range illustration (black solid line) showing the contribution of fast residual atmospheric errors (red dashed line) and quasi-static aberrations (blue dotted line). The simulation includes both photon and speckle noise, and the quasi-static aberrations are assumed to follow a $f^{-3}$ power law with a 600s lifetime.}
 \label{Fig:DRspeckleQS}
\end{figure}
\subsection{Discussion on the effects of speckle reduction techniques}
The noise budgets expressed in Eq.\ref{Eq:VarQS2} and Eq.\ref{Eq:VarQS3} enable a general understanding of high contrast imaging for exoplanet detection. 
In the case of direct imaging without coronagraph, the main limitation is set by speckle pinning which appears as an amplification of the variance levels at the position of the maxima of the diffracted pattern. In the absence of static aberrations, pinning happens at the top of the Airy rings. 
In the presence of static aberrations and even with a perfect coronagraph, residual pinning exists at the position of the maxima of the coronagraphic PSF corresponding to the propagated static aberrations through the coronagraph. For example, in the case of low-order static aberrations described by Zernike polynomials, the location of the maxima of the PSF (direct or coronagraphic) dictates where pinning happens. These locations are no longer at the top of the perfect Airy pattern in this case. 
Quasi-static aberrations also add to the problem by creating additional pinning terms (Eq.\ref{Eq:sigmaI}).
This phenomenon was observed at the AEOS telescope with the Lyot project coronagraph, where pinned speckles have been associated with dead actuators on the deformable mirror \citep{SOP06}, or AO waffle mode \citep{MSP05}.

In reality, coronagraphs are not perfect and suffer from various sources of errors, including chromaticity, manufacturing imperfections, alignment errors, etc. 
The cancellation of speckle pinning in the final error budget provides a constraint on the coronagraph performance. This can be achieved based on the comparison between the terms $I_c$ and $I_s$ described above.

With current state of the art AO systems and coronagraphs alone, it is clear that the residual noise terms (even if the pinning term is well controlled) would not enable the detection of giant planets with typical contrasts of $10^{-7}$ at 0.5 arcsec, or Earth-like planets with contrast of $10^{-10}$ at 0.1 arcsec.
Additional speckle reduction techniques have been developed \citep{BT06,GKV06} to improve the contrast performance further in combination with wavefront control and coronagraphs. 
These techniques affect the term $I_s$ by modifying the pupil aberrations to create a dark zone in the search region of the field.
In the case of ground-based high contrast imagers, for example with the GPI speckle calibration system \citep{KBR06}, the speckle reduction will only affect the $I_{s2}$ term, as the sensing involved in the technique is slow. Some speckle reduction techniques may also affect the residual speckle pinning in some region of the field, as they create a static wavefront aberration which creates a particularly asymmetric PSF presenting a dark zone \citep{CKH06,SWT06}.

The combination of coronagraphy and speckle nulling will therefore remove most of the terms contributing to the noise variance $\sigma_I^2$.
The limit for the association of coronagraphy and speckle nulling is set by the level of residual fast atmospheric aberrations.
Also, by decreasing the term $I_{s2}$, the speckle reduction system puts a stronger constraint on the coronagraph, whose performance must suppress pinned speckles to a level below other speckles in the dark region produced by speckle nulling.

Finally, the remaining parts of the speckle noise which have not been removed by a coronagraph or by the speckle reduction system, can be removed further using post-processing differential methods, combining multi-wavelength information from dual-band imaging, or an integral field spectrograph \cite{MDR00,SF02,MDN05}, or using differential rotation \cite{MLD06}. These differential methods affect all the residual speckle variance contributions (pinning and halo) in the same way. 
The combination of these three stages (coronagraph, speckle reduction and speckle processing) is necessary to reach the necessary dynamic range regime for exoplanet imaging. This approach has been chosen for the design of GPI \citep{MGP06}, and have been implemented in the laboratory by \citet{TT07}.

\section{Conclusion}

This paper develops theoretical insight in high contrast imaging, using a statistical model to understand the performance limitations of coronagraphic experiments. 
We discuss the application of the statistical model \citep{AS04} to the case of coronagraphic images, and show that the statistical model is valid outside of the focal plane mask occulting area. We show some limitations of the statistical model at the central point of the image, and a modification of the model is proposed by \citet{SF07} to solve this problem. 

It has been confirmed by several observations that the main limitation for high contrast imaging is due to the presence of quasi-static speckles. These do not average out over time, and are difficult to calibrate. We presented the first theoretical attempt to understand the effect of quasi-static speckles and their interaction with other aberrations (static and residual atmospheric). We obtained an analytical expression for the variance of the intensity, which includes speckle and photon noise in the presence of static, quasi-static and rapidly-varying aberrations. This result enables the use of a semi-analytical method, which has been compared successfully with purely numerical simulations. This semi-analytical method enables fast simulations and the possibility to compare and combine easily various types of parameters for instrument design studies. It also enables a breakdown of the noise into specific contributions (pinning, halo, speckle, photon). This method can be used to model real observations and extract information on relative noise contributions from various sources (residual atmospheric noise, quasi-static noise). 

The model also provides an understanding of the speckle pinning effect in high Strehl images, where bright speckles appear at the location of the diffraction rings. This effect, explained by a coherent amplification of the speckle noise, and can be cancelled by a coronagraph. The theory provides insight on the required level of performance for the coronagraph to effectively cancel speckle pinning. In the presence of quasi-static aberrations, we showed that pinning also exists between the perfect part of the wave and quasi-static aberrations, but is weighted by the ratio of the lifetimes (which can be a large number of the order of a hundred or more). This puts strong requirements on the level of static and quasi-static aberrations and may explain the results currently obtained with the Lyot project coronagraph, the first instrument of its kind to use an extreme AO system. Detail modeling of the Lyot project data, and specific simulations for the future generation of high contrast imager (GPI), will be done in the future.

This model was developed for the monochromatic case. Generalization to wide band is not straightforward, as the speckle noise is highly correlated with wavelength, which makes this generalization difficult, but can be used for multi-wavelength speckle subtractions. Also, a generalization of the model to the case of subtraction residuals in multi-wavelength imaging would be particularly interesting and is a topic for future study.

Although this presentation is focused on the ground-based case, the same formalism can be directly used for space-based observations, with similar results, but different orders of magnitudes (in particular for the speckle lifetimes). An identical expression of the variance of the noise was established independently by \citet{SMG05} for the TPF error budget. 

Once a coronagraph has removed the speckle pinning contribution, the dynamic range can be improved further by use of speckle reduction techniques (speckle nulling) or speckle subtraction (ADI, multi-wavelength or polarization). The effect of speckle nulling on the noise budget is now well understood from the expression of the variance. The model can be used to draw the requirements on the relative performance of the coronagraph and speckle reduction system. 
This theory brings an understanding of the effects on the dynamic range of the various elements in a high contrast instrument, including wavefront control, coronagraphy, speckle reduction and differential calibration techniques.
 All these stages are crucial and necessary to reach high contrast imaging, and they are going to be implemented in the next generation of high contrast imagers, currently in development.

\acknowledgements
R\'emi Soummer is supported by a Michelson Postdoctoral Fellowship, under contract to the Jet Propulsion Laboratory (JPL) funded by NASA. JPL is managed for NASA by the California Institute of Technology.
This work is based upon work partially supported by the National Science Fundation under Grant number No. AST-0215793 and AST-0334916 and has also been supported in part by the National Science Foundation Science and Technology Center for Adaptive Optics, managed by the University of California at Santa Cruz under cooperative agreement AST 98-76783.
The authors would like to thank Julian Christou, Szymon Gladysz, James P. Lloyd, Louis Lyons, Bruce Macintosh, Russell Makidon, Christian Marois and Anand Sivaramakrishnan for interesting discussions.



\begin{thebibliography}{101}

\bibitem[{{Aime}(2005{\natexlab{a}})}]{aime05b}
{Aime}, C. 2005{\natexlab{a}}, \pasp, 117, 1012

\bibitem[{{Aime}(2005{\natexlab{b}})}]{Aime05a}
---. 2005{\natexlab{b}}, \aap, 434, 785

\bibitem[{{Aime} \& {Soummer}(2004{\natexlab{a}})}]{AS04a}
{Aime}, C. \& {Soummer}, R. 2004{\natexlab{a}}, in EAS Publications Series, ed.
  C.~{Aime} \& R.~{Soummer}, 89--101

\bibitem[{{Aime} \& {Soummer}(2004{\natexlab{b}})}]{AS04}
{Aime}, C. \& {Soummer}, R. 2004{\natexlab{b}}, \apjl, 612, L85

\bibitem[{{Aime} {et~al.}(2002){Aime}, {Soummer}, \& {Ferrari}}]{ASF02}
{Aime}, C., {Soummer}, R., \& {Ferrari}, A. 2002, A\&A, 389, 334

\bibitem[{{Angel}(1994)}]{A94}
{Angel}, J.~R.~P. 1994, \nat, 368, 203

\bibitem[{{Baraffe} {et~al.}(2003){Baraffe}, {Chabrier}, {Barman}, {Allard}, \&
  {Hauschildt}}]{BCB03}
{Baraffe}, I., {Chabrier}, G., {Barman}, T.~S., {Allard}, F., \& {Hauschildt},
  P.~H. 2003, \aap, 402, 701

\bibitem[{{Beuzit} {et~al.}(2005){Beuzit}, {Mouillet}, {Dohlen}, \&
  {Puget}}]{BMD05}
{Beuzit}, J.-L., {Mouillet}, D., {Dohlen}, K., \& {Puget}, P. 2005, in
  SF2A-2005: Semaine de l'Astrophysique Francaise, ed. F.~{Casoli},
  T.~{Contini}, J.~M. {Hameury}, \& L.~{Pagani}, 215--+

\bibitem[{{Beuzit} {et~al.}(1997){Beuzit}, {Mouillet}, {Lagrange}, \&
  {Paufique}}]{BML97}
{Beuzit}, J.-L., {Mouillet}, D., {Lagrange}, A.-M., \& {Paufique}, J. 1997,
  A\&AS, 125, 175

\bibitem[{{Bloemhof}(2003{\natexlab{a}})}]{B03}
{Bloemhof}, E.~E. 2003{\natexlab{a}}, ApJL, 582, L59

\bibitem[{{Bloemhof}(2003{\natexlab{b}})}]{Blo03}
---. 2003{\natexlab{b}}, ApJL, 582, L59

\bibitem[{{Bloemhof}(2004{\natexlab{a}})}]{Blo04b}
---. 2004{\natexlab{a}}, \apjl, 610, L69

\bibitem[{{Bloemhof}(2004{\natexlab{b}})}]{Blo04}
---. 2004{\natexlab{b}}, Optics Letters, 29, 2333

\bibitem[{{Bloemhof} {et~al.}(2001){Bloemhof}, {Dekany}, {Troy}, \&
  {Oppenheimer}}]{BDT01}
{Bloemhof}, E.~E., {Dekany}, R.~G., {Troy}, M., \& {Oppenheimer}, B.~R. 2001,
  \apjl, 558, L71

\bibitem[{{Boccaletti}(2004)}]{B04}
{Boccaletti}, A. 2004, in EAS Publications Series, ed. C.~{Aime} \&
  R.~{Soummer}, 165--176

\bibitem[{{Boccaletti} {et~al.}(2003){Boccaletti}, {Chauvin}, {Lagrange}, \&
  {Marchis}}]{BCL03}
{Boccaletti}, A., {Chauvin}, G., {Lagrange}, A.-M., \& {Marchis}, F. 2003,
  \aap, 410, 283

\bibitem[{{Boccaletti} {et~al.}(2004){Boccaletti}, {Riaud}, {Baudoz},
  {Baudrand}, {Rouan}, {Gratadour}, {Lacombe}, \& {Lagrange}}]{BRB04}
{Boccaletti}, A., {Riaud}, P., {Baudoz}, P., {Baudrand}, J., {Rouan}, D.,
  {Gratadour}, D., {Lacombe}, F., \& {Lagrange}, A.-M. 2004, \pasp, 116, 1061

\bibitem[{{Bord{\'e}} \& {Traub}(2006)}]{BT06}
{Bord{\'e}}, P.~J. \& {Traub}, W.~A. 2006, \apj, 638, 488

\bibitem[{Brillinger(1981)}]{Bri81}
Brillinger, D. 1981, Time Series : Data Analysis and Theory (McGraw-Hill)

\bibitem[{{Brown}(2004{\natexlab{a}})}]{Brown04b}
{Brown}, R.~A. 2004{\natexlab{a}}, \apj, 610, 1079

\bibitem[{{Brown}(2004{\natexlab{b}})}]{Brown04a}
---. 2004{\natexlab{b}}, \apj, 607, 1003

\bibitem[{{Brown}(2005)}]{Brown05}
---. 2005, \apj, 624, 1010

\bibitem[{{Burrows} {et~al.}(2004){Burrows}, {Sudarsky}, \& {Hubeny}}]{BSH04}
{Burrows}, A., {Sudarsky}, D., \& {Hubeny}, I. 2004, \apj, 609, 407

\bibitem[{{Cagigal} \& {Canales}(1998)}]{CC98}
{Cagigal}, M.~P. \& {Canales}, V.~F. 1998, Optics Letters, 23, 1072

\bibitem[{{Cagigal} \& {Canales}(2000)}]{CC00}
---. 2000, JOSA, 17, 1312

\bibitem[{{Canales} \& {Cagigal}(1999)}]{CC99}
{Canales}, V.~F. \& {Cagigal}, M.~P. 1999, AO, 38, 766

\bibitem[{{Canales} \& {Cagigal}(2001)}]{CC01}
---. 2001, Optics Letters, 26, 737

\bibitem[{{Cavarroc} {et~al.}(2006){Cavarroc}, {Boccaletti}, {Baudoz}, {Fusco},
  \& {Rouan}}]{CBB06}
{Cavarroc}, C., {Boccaletti}, A., {Baudoz}, P., {Fusco}, T., \& {Rouan}, D.
  2006, \aap, 447, 397

\bibitem[{{Chabrier} \& {Baraffe}(2000)}]{CB00}
{Chabrier}, G. \& {Baraffe}, I. 2000, \araa, 38, 337

\bibitem[{{Christou} {et~al.}(2006){Christou}, {Gladysz}, {Redfern},
  {Bradford}, \& {Roberts}}]{CGR06}
{Christou}, J.~C., {Gladysz}, S., {Redfern}, M., {Bradford}, L.~W., \&
  {Roberts}, L.~C.~J. 2006, in Proc AMOS technical conference, Maui, Hawaii,
  10-14 September 2006, 528--537

\bibitem[{{Codona} {et~al.}(2006){Codona}, {Kenworthy}, {Hinz}, {Angel}, \&
  {Woolf}}]{CKH06}
{Codona}, J.~L., {Kenworthy}, M.~A., {Hinz}, P.~M., {Angel}, J.~R.~P., \&
  {Woolf}, N.~J. 2006, in Ground-based and Airborne Instrumentation for
  Astronomy. Edited by McLean, Ian S.; Iye, Masanori. Proceedings of the SPIE,
  Volume 6269, pp. (2006).

\bibitem[{{Digby} {et~al.}(2006){Digby}, {Hinkley}, {Oppenheimer},
  {Sivaramakrishnan}, {Lloyd}, {Perrin}, {Roberts}, {Soummer}, {Brenner},
  {Makidon}, {Shara}, {Kuhn}, {Graham}, {Kalas}, \& {Newburgh}}]{DHO06}
{Digby}, A.~P., {Hinkley}, S., {Oppenheimer}, B.~R., {Sivaramakrishnan}, A.,
  {Lloyd}, J.~P., {Perrin}, M.~D., {Roberts}, L.~C.~J., {Soummer}, R.,
  {Brenner}, D., {Makidon}, R.~B., {Shara}, M., {Kuhn}, J., {Graham}, J.~R.,
  {Kalas}, P.~G., \& {Newburgh}, L. 2006, Submited to ApJ

\bibitem[{{Ferrari}(2006)}]{F06}
{Ferrari}, A. 2006, in EAS Publications Series, 85--101

\bibitem[{{Ferrari} {et~al.}(2006){Ferrari}, {Carbillet}, {Serradel}, {Aime},
  \& {Soummer}}]{FCS06}
{Ferrari}, A., {Carbillet}, M., {Serradel}, E., {Aime}, C., \& {Soummer}, R.
  2006, in Proceedings IAU Colloquium No. 200, ed. C.~{Aime} \& J.~{Vakili},
  F.~{Darcourt}

\bibitem[{{Fitzgerald} \& {Graham}(2006)}]{FG06}
{Fitzgerald}, M.~P. \& {Graham}, J.~R. 2006, \apj, 637, 541

\bibitem[{{Foo} {et~al.}(2005){Foo}, {Palacios}, \& {Swartzlander}}]{FPS05}
{Foo}, G., {Palacios}, D.~M., \& {Swartzlander}, Jr., G.~A. 2005, Optics
  Letters, 30, 3308

\bibitem[{{Fusco} \& {Conan}(2004)}]{FC04}
{Fusco}, T. \& {Conan}, J.-M. 2004, Journal of the Optical Society of America
  A, 21, 1277

\bibitem[{{Fusco} {et~al.}(2005){Fusco}, {Rousset}, {Mouillet}, \&
  {Beuzit}}]{FRM05}
{Fusco}, T., {Rousset}, G., {Mouillet}, D., \& {Beuzit}, J.-L. 2005, in
  SF2A-2005: Semaine de l'Astrophysique Francaise, ed. F.~{Casoli},
  T.~{Contini}, J.~M. {Hameury}, \& L.~{Pagani}, 219--+

\bibitem[{{Give'on} {et~al.}(2006){Give'on}, {Kasdin}, {Vanderbei}, \&
  {Avitzour}}]{GKV06}
{Give'on}, A., {Kasdin}, N.~J., {Vanderbei}, R.~J., \& {Avitzour}, Y. 2006,
  JOSA A, 23

\bibitem[{{Gladysz}(2006)}]{G06}
{Gladysz}, S. 2006, PhD thesis, National University of Ireland, Galway

\bibitem[{{Gladysz} {et~al.}(2006){Gladysz}, {Christou}, \& {Redfern}}]{GCR06}
{Gladysz}, S., {Christou}, J.~C., \& {Redfern}, M. 2006, in Advances in
  Adaptive Optics II. Edited by Ellerbroek, Brent L.; Bonaccini Calia,
  Domenico. Proceedings of the SPIE, Volume 6272, pp. (2006).

\bibitem[{{Goodman}(1975)}]{Goo75}
{Goodman}, J. 1975, in topics in applied physics : laser speckle and related
  phenomena, Dainty Ed. (springer verlag berlin)

\bibitem[{{Goodman}(1996)}]{Goo96}
{Goodman}, J. 1996, Introduction to Fourier Optics (Mac Graw Hill)

\bibitem[{Goodman(2000)}]{Goo00}
Goodman, J.~W. 2000, Statistical Optics (Wiley Classics Library), wiley
  classics library ed. edn., Wiley classics library (New York:
  Wiley-Interscience), 576

\bibitem[{Goodman(2006)}]{Goo06}
---. 2006, Speckle Phenomena in Optics (Englewood, Colo.: Roberts and Company
  Publishers), 384

\bibitem[{{Guyon}(2005)}]{G05}
{Guyon}, O. 2005, \apj, 629, 592

\bibitem[{Hardy(1998)}]{Hardy98}
Hardy, J.~W. 1998, Adaptive Optics for Astronomical Telescopes (Oxford
  University Press, USA), 448

\bibitem[{{Hayward} {et~al.}(2001){Hayward}, {Brandl}, {Pirger}, {Blacken},
  {Gull}, {Schoenwald}, \& {Houck}}]{HBP01}
{Hayward}, T.~L., {Brandl}, B., {Pirger}, B., {Blacken}, C., {Gull}, G.~E.,
  {Schoenwald}, J., \& {Houck}, J.~R. 2001, \pasp, 113, 105

\bibitem[{{Hinkley} {et~al.}(2006){Hinkley}, {Oppenheimer}, {Soummer},
  {Sivaramakrishnan}, {Roberts}, {Kuhn}, {Makidon}, {Perrin}, {Lloyd},
  {Kratter}, \& {Brenner}}]{HOS06}
{Hinkley}, S., {Oppenheimer}, B.~R., {Soummer}, R., {Sivaramakrishnan}, A.,
  {Roberts}, L.~C.~J., {Kuhn}, J., {Makidon}, R.~B., {Perrin}, M.~D., {Lloyd},
  J.~P., {Kratter}, K., \& {Brenner}, D. 2006, Submited to ApJ

\bibitem[{{Jacquinot} \& {Roizen-Dossier}(1964)}]{JR64}
{Jacquinot}, P. \& {Roizen-Dossier}, B. 1964, Progress in Optics, Vol.~3
  ({Wolf}, E.)

\bibitem[{{Johnson} {et~al.}(1995){Johnson}, {Kotz}, \& {Balakrishnan}}]{JKB95}
{Johnson}, N., {Kotz}, S., \& {Balakrishnan}, N. 1995, Continuous Univariate
  Distributions, Vol.~2 (John Wiley \& Sons Inc.)

\bibitem[{{Jolissaint} {et~al.}(2006){Jolissaint}, {V{\'e}ran}, \&
  {Conan}}]{JVC06}
{Jolissaint}, L., {V{\'e}ran}, J.-P., \& {Conan}, R. 2006, Journal of the
  Optical Society of America A, 23, 382

\bibitem[{{Kasdin} {et~al.}(2003){Kasdin}, {Vanderbei}, {Spergel}, \&
  {Littman}}]{KVS03}
{Kasdin}, N.~J., {Vanderbei}, R.~J., {Spergel}, D.~N., \& {Littman}, M.~G.
  2003, \apj, 582, 1147

\bibitem[{{Kuchner} \& {Spergel}(2003)}]{KS03}
{Kuchner}, M.~J. \& {Spergel}, D.~N. 2003, \apj, 594, 617

\bibitem[{{Kuchner} \& {Traub}(2002)}]{KT02}
{Kuchner}, M.~J. \& {Traub}, W.~A. 2002, ApJ, 570, 900

\bibitem[{{Lloyd} {et~al.}(2006){Lloyd}, {Martinache}, {Ireland}, {Monnier},
  {Pravdo}, {Shaklan}, \& {Tuthill}}]{LMI06}
{Lloyd}, J.~P., {Martinache}, F., {Ireland}, M.~J., {Monnier}, J.~D., {Pravdo},
  S.~H., {Shaklan}, S.~B., \& {Tuthill}, P.~G. 2006, \apjl, 650, L131

\bibitem[{{Lyot}(1939)}]{L39}
{Lyot}, B. 1939, \mnras, 99, 580

\bibitem[{{Macintosh} {et~al.}(2006){Macintosh}, {Graham}, {Palmer}, {Doyon},
  {Gavel}, {Larkin}, {Oppenheimer}, {Saddlemyer}, {Wallace}, {Bauman}, {Evans},
  {Erikson}, {Morzinski}, {Phillion}, {Poyneer}, {Sivaramakrishnan}, {Soummer},
  {Thibault}, \& {Veran}}]{MGP06}
{Macintosh}, B., {Graham}, J., {Palmer}, D., {Doyon}, R., {Gavel}, D.,
  {Larkin}, J., {Oppenheimer}, B., {Saddlemyer}, L., {Wallace}, J.~K.,
  {Bauman}, B., {Evans}, J., {Erikson}, D., {Morzinski}, K., {Phillion}, D.,
  {Poyneer}, L., {Sivaramakrishnan}, A., {Soummer}, R., {Thibault}, S., \&
  {Veran}, J.-P. 2006, in Advances in Adaptive Optics II. Edited by Ellerbroek,
  Brent L.; Bonaccini Calia, Domenico. Proceedings of the SPIE, Volume 6272,
  pp. (2006).

\bibitem[{{Macintosh} {et~al.}(2004){Macintosh}, {Bauman}, {Wilhelmsen Evans},
  {Graham}, {Lockwood}, {Poyneer}, {Dillon}, {Gavel}, {Green}, {Lloyd},
  {Makidon}, {Olivier}, {Palmer}, {Perrin}, {Severson}, {Sheinis},
  {Sivaramakrishnan}, {Sommargren}, {Soummer}, {Troy}, {Wallace}, \&
  {Wishnow}}]{MBW04}
{Macintosh}, B.~A., {Bauman}, B., {Wilhelmsen Evans}, J., {Graham}, J.~R.,
  {Lockwood}, C., {Poyneer}, L., {Dillon}, D., {Gavel}, D.~T., {Green}, J.~J.,
  {Lloyd}, J.~P., {Makidon}, R.~B., {Olivier}, S., {Palmer}, D., {Perrin},
  M.~D., {Severson}, S., {Sheinis}, A.~I., {Sivaramakrishnan}, A.,
  {Sommargren}, G., {Soummer}, R., {Troy}, M., {Wallace}, J.~K., \& {Wishnow},
  E. 2004, in Advancements in Adaptive Optics. Edited by Domenico B. Calia,
  Brent L. Ellerbroek, and Roberto Ragazzoni. Proceedings of the SPIE, Volume
  5490, pp. 359-369 (2004)., ed. D.~{Bonaccini Calia}, B.~L. {Ellerbroek}, \&
  R.~{Ragazzoni}, 359--369

\bibitem[{{Makidon} {et~al.}(2005){Makidon}, {Sivaramakrishnan}, {Perrin},
  {Roberts}, {Oppenheimer}, {Soummer}, \& {Graham}}]{MSP05}
{Makidon}, R.~B., {Sivaramakrishnan}, A., {Perrin}, M.~D., {Roberts}, L.~C.,
  {Oppenheimer}, B.~R., {Soummer}, R., \& {Graham}, J.~R. 2005, \pasp, 117, 831

\bibitem[{{Malbet} {et~al.}(1995){Malbet}, {Yu}, \& {Shao}}]{MYS95}
{Malbet}, F., {Yu}, J.~W., \& {Shao}, M. 1995, \pasp, 107, 386

\bibitem[{{Marois} {et~al.}(2000){Marois}, {Doyon}, {Racine}, \&
  {Nadeau}}]{MDR00}
{Marois}, C., {Doyon}, R.~., {Racine}, R.~., \& {Nadeau}, D. 2000, PASP, 112,
  91

\bibitem[{{Marois} {et~al.}(2005){Marois}, {Doyon}, {Nadeau}, {Racine},
  {Riopel}, {Vall{\'e}e}, \& {Lafreni{\`e}re}}]{MDN05}
{Marois}, C., {Doyon}, R., {Nadeau}, D., {Racine}, R., {Riopel}, M.,
  {Vall{\'e}e}, P., \& {Lafreni{\`e}re}, D. 2005, \pasp, 117, 745

\bibitem[{{Marois} {et~al.}(2003){Marois}, {Doyon}, {Nadeau}, {Racine}, \&
  {Walker}}]{MDN03}
{Marois}, C., {Doyon}, R., {Nadeau}, D., {Racine}, R., \& {Walker}, G.~A.~H.
  2003, in EAS Publications Series, ed. C.~{Aime} \& R.~{Soummer}, 233--243

\bibitem[{{Marois} {et~al.}(2006{\natexlab{a}}){Marois}, {Lafreni{\`e}re},
  {Doyon}, {Macintosh}, \& {Nadeau}}]{MLD06}
{Marois}, C., {Lafreni{\`e}re}, D., {Doyon}, R., {Macintosh}, B., \& {Nadeau},
  D. 2006{\natexlab{a}}, \apj, 641, 556

\bibitem[{{Marois} {et~al.}(2006{\natexlab{b}}){Marois}, {Lafreni{\`e}re},
  {Macintosh}, \& {Doyon}}]{MLM06}
{Marois}, C., {Lafreni{\`e}re}, D., {Macintosh}, B., \& {Doyon}, R.
  2006{\natexlab{b}}, \apj, 647, 612

\bibitem[{{Marois} {et~al.}(2006{\natexlab{c}}){Marois}, {Phillion}, \&
  {Macintosh}}]{MPM06}
{Marois}, C., {Phillion}, D.~W., \& {Macintosh}, B. 2006{\natexlab{c}}, in
  Ground-based and Airborne Instrumentation for Astronomy. Edited by McLean,
  Ian S.; Iye, Masanori. Proceedings of the SPIE, Volume 6269, pp. (2006).

\bibitem[{{Masciadri} {et~al.}(2005){Masciadri}, {Mundt}, {Henning}, {Alvarez},
  \& {Barrado y Navascu{\'e}s}}]{MMH05}
{Masciadri}, E., {Mundt}, R., {Henning}, T., {Alvarez}, C., \& {Barrado y
  Navascu{\'e}s}, D. 2005, \apj, 625, 1004

\bibitem[{{Mawet} {et~al.}(2005){Mawet}, {Riaud}, {Absil}, \& {Surdej}}]{MRA05}
{Mawet}, D., {Riaud}, P., {Absil}, O., \& {Surdej}, J. 2005, \apj, 633, 1191

\bibitem[{McCullagh(1987)}]{MC87}
McCullagh, P. 1987, Tensor methods in statistics, Monographs on statistics and
  applied probability (Wiley, New York)

\bibitem[{{Michel} \& {Ferrari}(2003)}]{MF03}
{Michel}, O. \& {Ferrari}, A. 2003, in EAS Publications Series, ed. C.~{Aime}
  \& R.~{Soummer}, 129--146

\bibitem[{{Moffat}(1969)}]{Mof69}
{Moffat}, A.~F.~J. 1969, A\&A, 3, 455

\bibitem[{{Nisenson} \& {Papaliolios}(2001)}]{NP01}
{Nisenson}, P. \& {Papaliolios}, C. 2001, ApJ Letters, 549

\bibitem[{{Oppenheimer} {et~al.}(2004){Oppenheimer}, {Digby}, {Newburgh},
  {Brenner}, {Shara}, {Mey}, {Mandeville}, {Makidon}, {Sivaramakrishnan},
  {Soummer}, {Graham}, {Kalas}, {Perrin}, {Roberts}, {Kuhn}, {Whitman}, \&
  {Lloyd}}]{ODN04}
{Oppenheimer}, B.~R., {Digby}, A.~P., {Newburgh}, L., {Brenner}, D., {Shara},
  M., {Mey}, J., {Mandeville}, C., {Makidon}, R.~B., {Sivaramakrishnan}, A.,
  {Soummer}, R., {Graham}, J.~R., {Kalas}, P., {Perrin}, M.~D., {Roberts},
  L.~C., {Kuhn}, J.~R., {Whitman}, K., \& {Lloyd}, J.~P. 2004, in Advancements
  in Adaptive Optics. Edited by Domenico B. Calia, Brent L. Ellerbroek, and
  Roberto Ragazzoni. Proceedings of the SPIE, Volume 5490, pp. 433-442 (2004).,
  ed. D.~{Bonaccini Calia}, B.~L. {Ellerbroek}, \& R.~{Ragazzoni}, 433--442

\bibitem[{{Oppenheimer} {et~al.}(2001){Oppenheimer}, {Golimowski}, {Kulkarni},
  {Matthews}, {Nakajima}, {Creech-Eakman}, \& {Durrance}}]{OGK01}
{Oppenheimer}, B.~R., {Golimowski}, D.~A., {Kulkarni}, S.~R., {Matthews}, K.,
  {Nakajima}, T., {Creech-Eakman}, M., \& {Durrance}, S.~T. 2001, \aj, 121,
  2189

\bibitem[{{Perrin} {et~al.}(2003){Perrin}, {Sivaramakrishnan}, {Makidon},
  {Oppenheimer}, \& {Graham}}]{PSM03}
{Perrin}, M.~D., {Sivaramakrishnan}, A., {Makidon}, R., {Oppenheimer}, B.~R.,
  \& {Graham}, J.~R. 2003, ApJ, 596, 702

\bibitem[{{Poyneer} \& {Macintosh}(2004)}]{PM04}
{Poyneer}, L.~A. \& {Macintosh}, B. 2004, Journal of the Optical Society of
  America A, vol.~21, Issue 5, pp.810-819, 21, 810

\bibitem[{{Racine}(1996)}]{Rac96}
{Racine}, R. 1996, \pasp, 108, 699

\bibitem[{{Racine} {et~al.}(1999){Racine}, {Walker}, {Nadeau}, {Doyon}, \&
  {Marois}}]{RWN99}
{Racine}, R., {Walker}, G.~A.~H., {Nadeau}, D., {Doyon}, R., \& {Marois}, C.
  1999, PASP, 111, 587

\bibitem[{{Roddier}(1981)}]{R81}
{Roddier}, F. 1981, Prog.~Optics, Volume 19, p.~281-376, 19, 281

\bibitem[{{Rouan} {et~al.}(2000){Rouan}, {Riaud}, {Boccaletti}, {Cl{\'e}net},
  \& {Labeyrie}}]{RRB00}
{Rouan}, D., {Riaud}, P., {Boccaletti}, A., {Cl{\'e}net}, Y., \& {Labeyrie}, A.
  2000, PASP, 112, 1479

\bibitem[{{Serabyn} {et~al.}(2006){Serabyn}, {Wallace}, {Troy}, {Mennesson},
  {Haguenauer}, {Gappinger}, \& {Bloemhof}}]{SWT06}
{Serabyn}, E., {Wallace}, J.~K., {Troy}, M., {Mennesson}, B., {Haguenauer}, P.,
  {Gappinger}, R.~O., \& {Bloemhof}, E.~E. 2006, in Advances in Adaptive Optics
  II. Edited by Ellerbroek, Brent L.; Bonaccini Calia, Domenico. Proceedings of
  the SPIE, Volume 6272, pp. (2006).

\bibitem[{{Shaklan} {et~al.}(2005){Shaklan}, {Marchen}, {Green}, \&
  {Lay}}]{SMG05}
{Shaklan}, S.~B., {Marchen}, L., {Green}, J.~J., \& {Lay}, O.~P. 2005, in
  Techniques and Instrumentation for Detection of Exoplanets II. Edited by
  Coulter, Daniel R. Proceedings of the SPIE, Volume 5905, pp. 110-121 (2005).,
  ed. D.~R. {Coulter}, 110--121

\bibitem[{{Sivaramakrishnan} {et~al.}(2003){Sivaramakrishnan}, {Hodge},
  {Makidon}, {Perrin}, {Lloyd}, {Bloemhof}, \& {Oppenheimer}}]{SHM03}
{Sivaramakrishnan}, A., {Hodge}, P.~E., {Makidon}, R.~B., {Perrin}, M.~D.,
  {Lloyd}, J.~P., {Bloemhof}, E.~E., \& {Oppenheimer}, B.~R. 2003, in
  High-Contrast Imaging for Exo-Planet Detection. Edited by Alfred B. Schultz.
  Proceedings of the SPIE, Volume 4860, pp. 161-170 (2003)., 161--170

\bibitem[{{Sivaramakrishnan} {et~al.}(2001){Sivaramakrishnan}, {Koresko},
  {Makidon}, {Berkefeld}, \& {Kuchner}}]{SKM01}
{Sivaramakrishnan}, A., {Koresko}, C.~D., {Makidon}, R.~B., {Berkefeld}, T., \&
  {Kuchner}, M.~J. 2001, \apj, 552, 397

\bibitem[{{Sivaramakrishnan} {et~al.}(2002){Sivaramakrishnan}, {Lloyd},
  {Hodge}, \& {Macintosh}}]{SLH02}
{Sivaramakrishnan}, A., {Lloyd}, J.~P., {Hodge}, P.~E., \& {Macintosh}, B.~A.
  2002, ApJL, 581, L59

\bibitem[{{Sivaramakrishnan} \& {Oppenheimer}(2006)}]{SO06}
{Sivaramakrishnan}, A. \& {Oppenheimer}, B.~R. 2006, \apj, 647, 620

\bibitem[{{Sivaramakrishnan} {et~al.}(2006){Sivaramakrishnan}, {Oppenheimer},
  {Perrin}, {Roberts}, {Makidon}, {Soummer}, {Digby}, {Bradford}, {Skinner},
  {Turner}, \& {Ten Brummelaar}}]{SOP06}
{Sivaramakrishnan}, A., {Oppenheimer}, B.~R., {Perrin}, M.~D., {Roberts},
  L.~C., {Makidon}, R.~B., {Soummer}, R., {Digby}, A.~P., {Bradford}, L.~W.,
  {Skinner}, M.~A., {Turner}, N.~H., \& {Ten Brummelaar}, T.~A. 2006, in IAU
  Colloq. 200: Direct Imaging of Exoplanets: Science {\&} Techniques, ed.
  C.~{Aime} \& F.~{Vakili}, 613--616

\bibitem[{{Sivaramakrishnan} {et~al.}(2007){Sivaramakrishnan}, {Soummer},
  {Oppenheimer}, \& {Pueyo}}]{SSO07}
{Sivaramakrishnan}, A., {Soummer}, R., {Oppenheimer}, B., \& {Pueyo}, L. 2007,
  Submited to ApJ

\bibitem[{{Sivaramakrishnan} {et~al.}(2005){Sivaramakrishnan}, {Soummer},
  {Sivaramakrishnan}, {Lloyd}, {Oppenheimer}, \& {Makidon}}]{SSS05}
{Sivaramakrishnan}, A., {Soummer}, R., {Sivaramakrishnan}, A.~V., {Lloyd},
  J.~P., {Oppenheimer}, B.~R., \& {Makidon}, R.~B. 2005, \apj, 634, 1416

\bibitem[{{Soummer}(2005)}]{S05}
{Soummer}, R. 2005, \apjl, 618, L161

\bibitem[{{Soummer} {et~al.}(2003{\natexlab{a}}){Soummer}, {Aime}, \&
  {Falloon}}]{SAF03}
{Soummer}, R., {Aime}, C., \& {Falloon}, P.~E. 2003{\natexlab{a}}, \aap, 397,
  1161

\bibitem[{{Soummer} {et~al.}(2003{\natexlab{b}}){Soummer}, {Aime}, {Ferrari},
  \& {Falloon}}]{SAFF03}
{Soummer}, R., {Aime}, C., {Ferrari}, A., \& {Falloon}, P.~E.
  2003{\natexlab{b}}, in High-Contrast Imaging for Exo-Planet Detection. Edited
  by Alfred B. Schultz. Proceedings of the SPIE, Volume 4860, pp. 211-220
  (2003)., ed. A.~B. {Schultz}, 211--220

\bibitem[{{Soummer} {et~al.}(2003{\natexlab{c}}){Soummer}, {Dohlen}, \&
  {Aime}}]{SDA03}
{Soummer}, R., {Dohlen}, K., \& {Aime}, C. 2003{\natexlab{c}}, \aap, 403, 369

\bibitem[{{Soummer} \& {Ferrari}(2007)}]{SF07}
{Soummer}, R. \& {Ferrari}, A. 2007, ApJL, in press

\bibitem[{{Soummer} {et~al.}(2006){Soummer}, {Oppenheimer}, {Hinkley},
  {Sivaramakrishnan}, {Makidon}, {Oppenheimer}, {Digby}, {Brenner}, {Kuhn},
  {Perrin}, {Roberts}, \& {Kratter}}]{SOH06}
{Soummer}, R., {Oppenheimer}, B.~R., {Hinkley}, S., {Sivaramakrishnan}, A.,
  {Makidon}, R.~B., {Oppenheimer}, B.~R., {Digby}, A.~P., {Brenner}, D.,
  {Kuhn}, J., {Perrin}, M.~D., {Roberts}, L.~C.~J., \& {Kratter}, K. 2006, in
  EAS Publications Series, ed. C.~{Aime} \& M.~{Carbillet}

\bibitem[{{Sparks} \& {Ford}(2002)}]{SF02}
{Sparks}, W.~B. \& {Ford}, H.~C. 2002, \apj, 578, 543

\bibitem[{{Trauger} \& {Traub}(2007)}]{TT07}
{Trauger}, J.~T. \& {Traub}, W.~A. 2007, Nature, 446, 771

\bibitem[{{Troy} {et~al.}(2000){Troy}, {Dekany}, {Brack}, {Oppenheimer},
  {Bloemhof}, {Trinh}, {Dekens}, {Shi}, {Hayward}, \& {Brandl}}]{TDB00}
{Troy}, M., {Dekany}, R.~G., {Brack}, G., {Oppenheimer}, B.~R., {Bloemhof},
  E.~E., {Trinh}, T., {Dekens}, F.~G., {Shi}, F., {Hayward}, T.~L., \&
  {Brandl}, B. 2000, in Proc. SPIE Vol. 4007, p. 31-40, Adaptive Optical
  Systems Technology, Peter L. Wizinowich; Ed., 31--40

\bibitem[{{Wallace} {et~al.}(2006){Wallace}, {Bartos}, {Rao}, {Samuele}, \&
  {Schmidtlin}}]{KBR06}
{Wallace}, J.~K., {Bartos}, R., {Rao}, S., {Samuele}, R., \& {Schmidtlin}, E.
  2006, in Advances in Adaptive Optics II. Edited by Ellerbroek, Brent L.;
  Bonaccini Calia, Domenico. Proceedings of the SPIE, Volume 6272, pp. (2006).

\bibitem[{{Wolfram}(1999)}]{W99}
{Wolfram}, S. 1999, The Mathematica book, Fourth Edition (Cambridge University
  Press)

\end{thebibliography}

\end{document}